\documentclass[iop]{emulateapj}
\usepackage{graphicx}
\usepackage{natbib}
\usepackage{amsmath}
\usepackage{amssymb}
\usepackage{comment}
\begin{document}

\title{A Young GMC Formed at the Interface of Two Colliding Supershells: \\Observations Meet Simulations}
\author{J. R. Dawson\altaffilmark{1,2}}
\author{E. Ntormousi\altaffilmark{3}}
\author{Y. Fukui\altaffilmark{4}}
\author{T. Hayakawa\altaffilmark{4}}
\author{K. Fierlinger\altaffilmark{5,6}}
\email{joanne.dawson@mq.edu.au}
\altaffiltext{1}{Department of Physics and Astronomy and MQ Research Centre in Astronomy, Astrophysics and Astrophotonics, Macquarie University, NSW 2109, Australia}
\altaffiltext{2}{Australia Telescope National Facility, CSIRO Astronomy and Space Science, PO Box 76, Epping, NSW 1710, Australia}
\altaffiltext{3}{Service dÕAstrophysique, CEA/DSM/IRFU Orme des Merisiers, Bat 709 Gif-sur-Yvette, 91191 France}
\altaffiltext{4}{Department of Physics and Astrophysics, Nagoya University, Chikusa-ku, Nagoya, Japan}
\altaffiltext{5}{University Observatory Munich, Scheinerstr. 1, D-81679 M\"unchen, Germany}
\altaffiltext{6}{Excellence Cluster Universe, Technische Universit\"at M\"unchen, Boltzmannstr. 2, D-85748, Garching, Germany}


\begin{abstract}
Dense, star-forming gas is believed to form at the stagnation points of large-scale ISM flows, but observational examples of this process in action are rare. We here present a giant molecular cloud (GMC) sandwiched between two colliding Milky Way supershells, which we argue shows strong evidence of having formed from material accumulated at the collision zone. Combining $^{12}$CO, $^{13}$CO and C$^{18}$O(J=1--0) data with new high-resolution, 3D hydrodynamical simulations of colliding supershells, we discuss the origin and nature of the GMC (G288.5+1.5), favoring a scenario in which the cloud was partially seeded by pre-existing denser material, but assembled into its current form by the action of the shells. This assembly includes the production of some new molecular gas. The GMC is well interpreted as non-self-gravitating, despite its high mass ($M_{\mathrm{H}_2}\sim1.7\times10^5~M_{\odot}$), and is likely pressure confined by the colliding flows, implying that self-gravity was not 
a necessary ingredient for 
its formation. Much of the molecular gas is relatively diffuse, and the cloud as a whole shows little evidence of star formation activity, supporting a scenario in which it is young and recently formed. Drip-like formations along its lower edge may be explained by fluid dynamical instabilities in the cooled gas. 

\end{abstract}


\section{Introduction}

The conversion of gas into stars begins with the 
formation of cold, dense clouds from the warmer, diffuse interstellar medium (ISM). In the modern, high-metallicity universe, cold/dense gas is generally synonymous with molecular gas, and the majority of star formation takes place deep within giant molecular clouds (GMCs). Understanding how these large agglomerations of molecular material form and evolve is therefore an important component of understanding the star formation process in galaxies. 

A key requirement of GMC formation is that a large quantity of what was previously diffuse, atomic material must end up concentrated into a small volume of space. This, together with increasing awareness of the ISM as a dynamic and turbulent medium, has led to the development of a paradigm in which molecular clouds are formed at the stagnation points of large-scale ISM flows 
\citep[e.g.][]{ballesteros99,hennebelle99,koyama00,audit05,vazquez06,Heitsch_06,inoue09}. Proposed astrophysical drivers of these flows include gravitational instabilities in galaxy disks \citep[e.g.][]{wada00,kim02,tasker09,bournaud10,elmegreen11}, spiral shocks \citep[e.g.][]{kim06,dobbs06,dobbs07}, and expanding supershells driven by correlated supernovae and stellar winds (e.g. \citealt{mccray87,hartmann01,ntormousi11}; see also review by \citealt{dawson13b}). 

Focussing on the role of stellar feedback, 
\citet{inutsuka14} have recently developed a ``bubble-dominated'' picture of molecular cloud formation. 
They propose a multi-generational model, 
in which GMCs are built up in the overlapping regions of Galactic supershells from cold H{\sc i}, which is formed readily by previous episodes of stellar feedback. This scenario is motivated in part by the difficulty of forming large quantities of molecular gas 
from pure warm neutral medium flows, particularly in the presence of magnetic fields, which 
oppose the creation of sufficiently dense material \citep[see also][]{inoue08,inoue09,inoue12}. Repeated episodes of shock-compression 
offer an attractive way to overcome these difficulties, by allowing clouds to be built up incrementally from pre-existing denser gas. 

In this picture, smaller molecular clouds may also be formed without the need for multiple compressive episodes, but only in isolated portions of shell walls where the magnetic field is aligned fortuitously with the flow direction. This is consistent with observational work, which indeed finds that molecular clouds are distributed sparsely throughout the walls of Galactic shells while cold H{\sc i} is more ubiquitous \citep{dawson11a}. It is also interesting to note 
the wealth of observational work detailing the association of Milky Way molecular clouds with expanding superstructures, including many well-known star-forming clouds in the local ISM (see review by \citealt{dawson13b} for a detailed listing). However, robust evidence for the \textit{formation} of such clouds due to feedback processes has remained rare.
 
We here report the case of a GMC, G288.5+1.5, sandwiched between two old, gently expanding supershells 
in the Carina Arm of the Milky Way. This massive ($M_{\mathrm{H}_2}\sim1.7\times10^5~M_{\odot}$) and relatively local ($D\sim2.6$ kpc) cloud is perhaps the best candidate discovered to-date for a 
GMC formed at the stagnation point of feedback-driven ISM flows -- in this case, the overlap region of two superbubbles. Furthermore, quantitative evidence already exists for molecular gas production in one of the shells 
\citep{dawson11a}, strongly suggesting that some or all of this GMC 
was indeed formed by the accumulation of matter between them.

This paper presents a detailed observational investigation of G288.5+1.5 and its surroundings, paired with new high-resolution, 3D hydrodynamical simulations of cold gas formation in colliding supershells. We begin in the following section by describing $^{12}$CO(J=1--0), $^{13}$CO(J=1--0) and C$^{18}$O(J=1--0) observations made with the NANTEN and Mopra telescopes, which form the observational backbone of this work. Section \ref{obsoverview} presents an overview of the observational and physical properties of the GMC and the surrounding region, summarizing the properties of the two shells, outlining evidence for the physical location of the molecular gas between them, and demonstrating its interaction with both objects. Section \ref{model} describes the numerical simulations, which provide a valuable model of the supershell collision process, and theoretical context in which the observational results are interpreted. We then draw on both the model and observational results to discuss the origin of the molecular gas in section \ref{origins}, 
examine the gravitational stability of the cloud in \ref{gravity}, 
and discuss possible instability structures in the molecular gas in section \ref{instabilities}. We finally summarize our conclusions in section \ref{conclusions}.


\section{Observations}
\label{observations}

\subsection{NANTEN CO Data}

The data used in this work were taken as part of the PhD thesis of \citet{matsunagaphd}, and are used here with kind permission. (We note that our analysis and scientific conclusions differ from that work.) Observations in the $^{12}$CO(J=1--0),  $^{13}$CO(J=1--0) and C$^{18}$O(J=1--0) lines (rest frequencies: 115.271 GHz, 110.201 GHz and 109.782 GHz) were made with the 4 m NANTEN telescope, located at the time in Las Campanas Observatory, Chile. The telescope half power beam width was $\sim2.6'$ at 115 GHz and $\sim2.7'$ at 110 and 109 GHz. $^{12}$CO(J=1--0) and  $^{13}$CO(J=1--0) observations were carried out by position switching between November 2001 and March 2002, with a 2 arcmin pointing grid and typical on-source integration times of $\sim40$ s and $\sim50$ s, respectively. CO$^{18}$(J=1--0) observations were targeted towards $^{13}$CO detections, and carried out in frequency switching mode with a frequency offset of 13 MHz and typical integration times of $\sim4$ minutes. 
The system temperature was calibrated with a hot load (paddle), and was typically $\sim220$ K at 115 GHz, and $\sim140$ K at 110 and 109 GHz (in a single side band), including the atmosphere towards the zenith. The 2048 channel acousto-optical spectrometer provided a total bandwidth of 40 MHz and an effective spectral resolution of 40 kHz, corresponding to a velocity coverage and resolution of 100 km s$^{-1}$ and 0.1 km s$^{-1}$. Oph East IRA ($\alpha_{B1950}=16^\mathrm{h} 29^\mathrm{m} 20^\mathrm{s}.9$, $\delta_{B1950}=-24\degr 22\arcmin 13\arcsec$) was observed as a standard calibrator source, with (main beam) radiation temperatures \citep{kutner81} of $T_R^*=15$, 10 and 4.4 K assumed for the $^{12}$CO, $^{13}$CO and C$^{18}$O lines respectively. The final RMS noise fluctuations in a 0.1 km s$^{-1}$ channel were $\sim0.4$, $\sim0.2$ and $\sim0.1$ K for the $^{12}$CO(J=1--0), $^{13}$CO(J=1--0) and CO$^{18}$O(J=1--0) lines. 
Note that all corrected main beam radiation temperatures, $T_R^*$, are referred to simply as ``brightness temperatures'' for the remainder of this paper. 

\begin{figure*}
\centering
\includegraphics[scale=1.05]{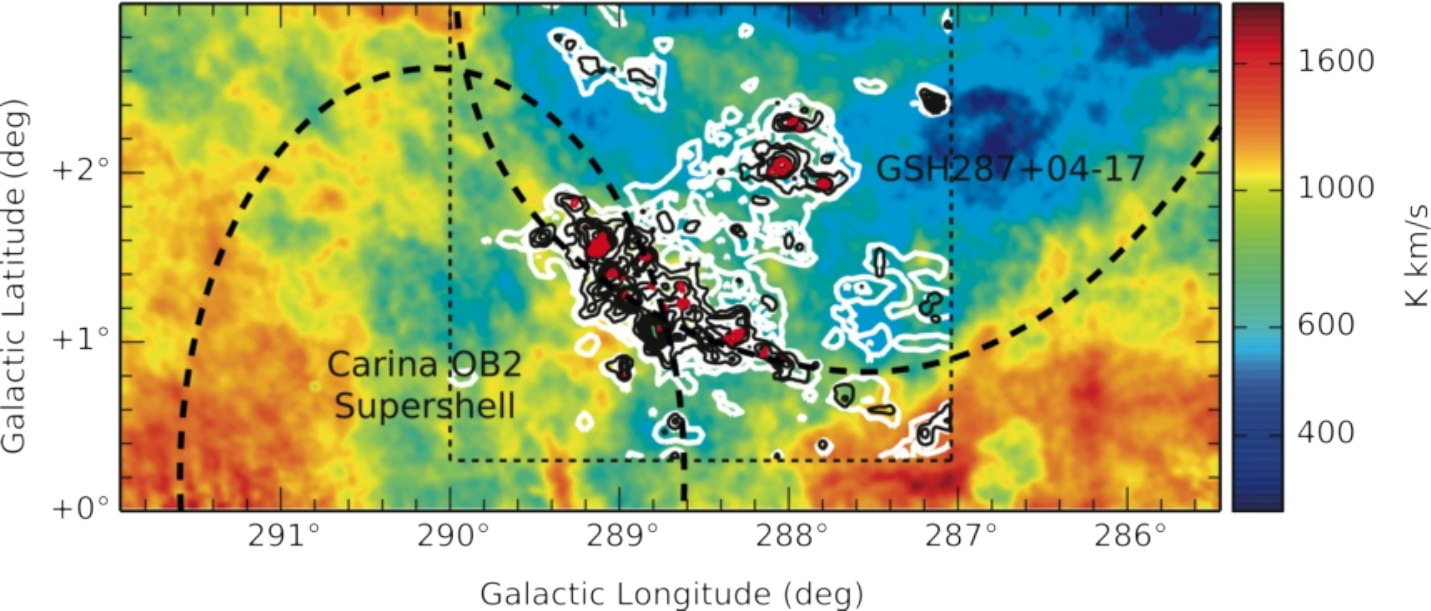}
\caption{Anatomy of the G288.5+1.5 region. The color image shows H{\sc i} data from \citet{dawson11a}, integrated over the velocity range of the GMC ($-33<v_{\mathrm{LSR}}<-11$ km s$^{-1}$). White contours are NANTEN $^{12}$CO(J=1--0) integrated over the same velocity range, beginning at 4.0 K km s$^{-1}$ and incremented every 5.0 K km s$^{-1}$ thereafter. Black contours are $^{13}$CO(J=1--0) velocity-integrated intensity, integrated over all velocity channels where $^{12}$CO was detected at the $3\sigma$ level, and drawn at intervals of 1.0 K km s$^{-1}$. Red contours are C$^{18}$O(J=1--0) velocity-integrated intensity, integrated over all velocity channels where $^{13}$CO was detected at the $3\sigma$ level, beginning at 0.3 K km s$^{-1}$ and incremented every 0.1 K km s$^{-1}$ thereafter. The thin dashed line marks the limits of the region observed in CO. The thick dashed lines mark the locations of the H{\sc i} supershells. For GSH 287+04--17 this line is a by-eye fit to the widest extent of the shell, which is delineated in part by the bright ridge of the GMC \citep[see][]{dawson08a}. For the Carina OB2 supershell the line is an ellipse approximating the dimensions of the H{\sc i} shell \citep{rizzo98} and should be considered only a very rough representation of its true shape.}
\label{gmc_lb}
\end{figure*}

\subsection{Mopra CO Data}

Higher resolution observations of two small sub-regions of the target GMC were observed in the $^{12}$CO(J=1--0),  $^{13}$CO(J=1--0) and CO$^{18}$O(J=1--0) lines in May/June 2014 using the Mopra telescope, near Coonabarabran, Australia. Observations were made in an on-the-fly (OTF) raster mapping mode, in which the telescope records data continuously while scanning across the sky. Each $7\arcmin\times7\arcmin$ map was observed at least twice in orthogonal scanning directions to minimize scanning artifacts. The scan speed was 3.5 arcsec\ s$^{-1}$, the sampling interval was 14\arcsec\ and the spacing between scan rows was 10\arcsec, fulfilling the minimum requirements for oversampling of the $\sim33\arcsec$ (FWHM) Mopra beam. For all sessions an off-source position was observed once per scan row. The pointing solution of the telescope was verified once every 90 minutes via observations of the SiO maser RW Vel ($\alpha_{J2000}=9^h20^m19.57^s$, $\delta_{J2000}=-49^{\circ}31\arcmin27.2\arcsec$), and corrections were applied for pointing errors of greater than 5$\arcsec$ in either azimuth or elevation. Paddle measurements were made every 15 minutes to calibrate the system temperature, which was also tracked in real-time with a noise diode. Typical values were $450$--$650$ K at 115 GHz and $250$--$350$K at 109 and 110 GHz. The backend was the MOPS digital filter bank, which simultaneously records dual polarization data for up to sixteen 137.5 MHz zoom bands positioned within an 8 GHz window. Zoom bands 
centered on the rest frequencies of the three lines 
each contained 4096 channels, providing a velocity resolution and coverage of 0.09 km s$^{-1}$ and $\sim360$ km s$^{-1}$ in all lines.

Bandpass calibration, baseline subtraction and calibration onto a $T_A^*$ scale were performed with the \textit{livedata} package. 
The spectra are then gridded into cubes using \textit{gridzilla}.
\footnote{Binaries and source code for \textit{livedata} and \textit{gridzilla} are available from http://www.atnf.csiro.au/computing/software/livedata.html.}
The data were weighted by the inverse of the system temperature, and convolved with a truncated Gaussian smoothing kernel with a FWHM of $60\arcsec$ and cutoff radius of $30\arcsec$ to improve the signal-to-noise ratio. This results in a final effective angular resolution of $55\arcsec$. 

The gridded data were converted to a main beam temperature scale ($T_R^*$) using a scaling factor determined by daily observations of the standard calibrator source Orion KL ($\alpha_{B1950}=5^h32^m47.5^s$, $\delta_{B1950}=-5^{\circ}24\arcmin21\arcsec$). The required scaling factors were $\eta=0.38\pm0.04$ for $^{12}$CO(J=1--0) and $0.50\pm0.04$ for $^{13}$CO(J=1--0), with a factor of $0.50$ also assumed for C$^{18}$O(J=1--0). These values are consistent with previous epochs. \cite[Further information on the scaling of Mopra data can be found in][and references within]{dawson11b}. The final cubes were binned in velocity to a channel width of 0.36 km s$^{-1}$. The RMS noise fluctuations in a 0.36 km s$^{-1}$ channel were $\sim0.33$, $\sim0.15$ and $\sim0.15$ K for the $^{12}$CO(J=1--0), $^{13}$CO(J=1--0) and CO$^{18}$O(J=1--0) lines, respectively. 


\section{Observational Overview} 
\label{obsoverview}

\subsection{Anatomy of the Region}

The Carina region contains two H{\sc i} supershells -- GSH 287+04--17 \citep[the `Carina Flare',][]{fukui99}, and the Carina OB supershell \citep{rizzo98}. The basic properties of these objects are summarized in table \ref{shellstable}, together with the references in which these were derived. Both are large ($R\sim100$ pc), gently expanding ($v_{\mathrm{exp}}\sim10$--20 km s$^{-1}$) H{\sc i} voids surrounded by denser swept-up shells. Both shells also have associated molecular gas. In the case of GSH 287+04--17, this molecular mass is approximately $\sim20\%$ of the total neutral gas mass of the system. (Mass estimates are not available for the Carina OB2 supershell.) The input energies required to form the shells have been roughly estimated as $\sim5\times10^{51}$ erg for both objects. Their distance estimates place them at $2.6\pm0.4$ and $2.9\pm0.9$ kpc, respectively -- coincident to within the uncertainties -- and they are likely to lie at least partially within the Carina Arm \citep{dawson08a}.

GSH 287+04--17 was studied in detail by \citet{dawson11b,dawson11a}. These authors make quantitative comparisons of the molecular gas fraction in the shell system, and compare it with that of the undisturbed ISM to argue that as much as half of the molecular mass in the system may have been formed as a direct result of the sweep-up and compression of material in the expanding supershell. The associated molecular cloud population is scattered throughout the H{\sc i} shell walls, and includes some moderately massive ($M_{\mathrm{H}_2}\sim10^4~M_{\odot}$) clouds at unusually high altitudes ($z\sim450$ pc), as well as the large GMC, closer to the Galactic Midplane, that is the subject of this paper. The molecular gas itself is not unusual -- the statistical properties of the population of clouds are indistinguishable from other Milky Way samples \citep{dawson08b}.  

\begin{figure}
\includegraphics[scale=0.57]{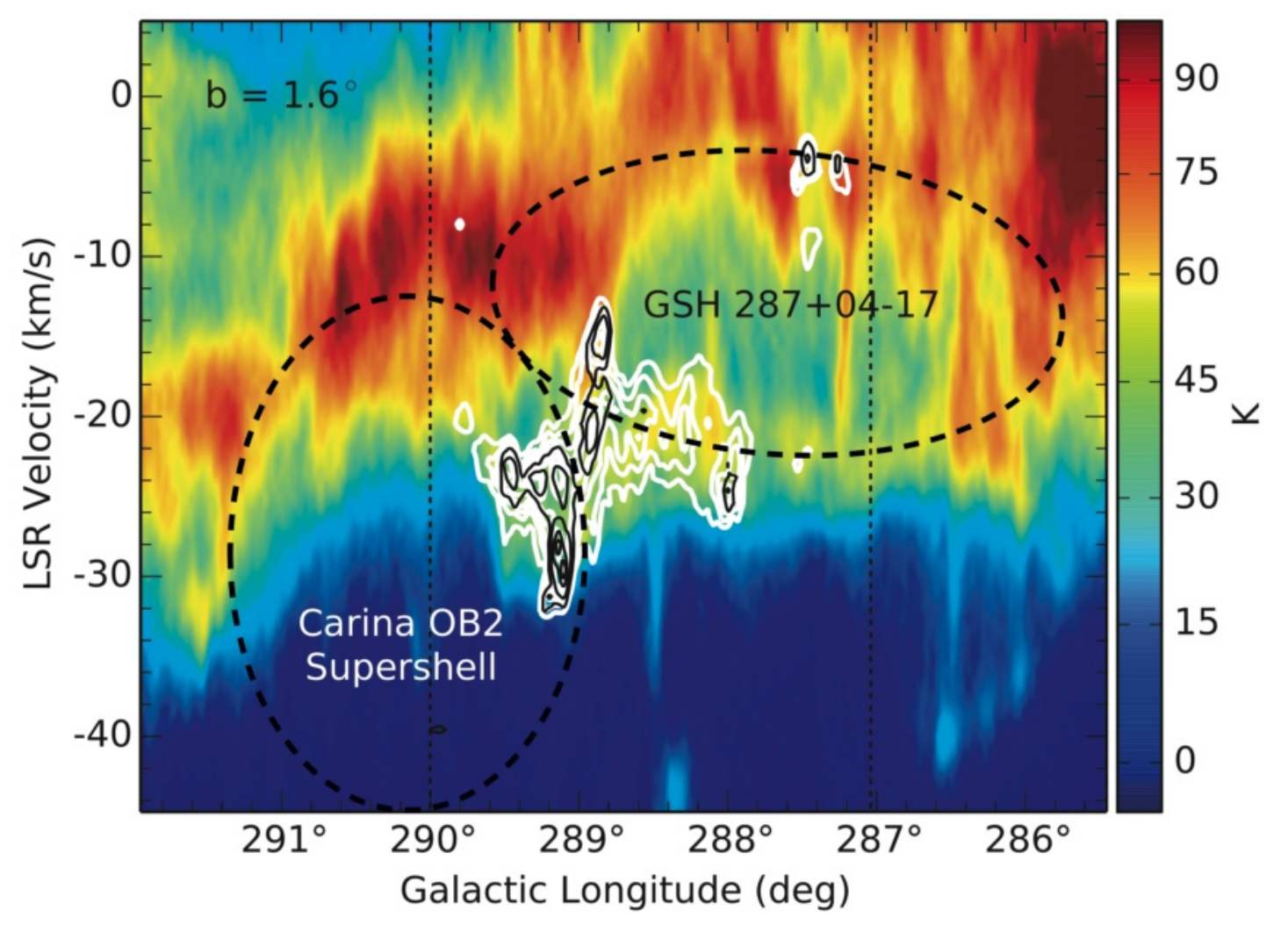}
\caption{Longitude-velocity plot of the G288.5+1.5 region, averaged over a latitude width of 10 arcmin, centered on $b=1.6^{\circ}$. The color image shows H{\sc i} and the white and black contours show $^{12}$CO(J=1--0) and $^{13}$CO(J=1--0). The dashed ellipses mark the expanding supershells. For GSH 287+04--17 the ellipse is a least squares fit to the H{\sc i} and CO intensity peaks taken directly from \citet{dawson08a}. For the Carina OB2 supershell, the ellipse is an approximation computed from the idealized dimensions, position and expansion velocity of the shell given in \citet{rizzo98}. Here, the the line-of-sight expansion velocity is scaled to reflect the offset between this latitude slice and the shell centroid ($b=0.2^{\circ}$), under the assumption that the depth of the shell along the line of sight is equal to its minor axis parallel to the Galactic Plane. }
\label{gmc_lv}
\end{figure}

The GMC itelf 
\citep[G288.5+1.5,][]{matsunagaphd} is located at exactly the position on the sky where the edges of two expanding H{\sc i} supershells intersect (see Figure \ref{gmc_lb}), with a bright ridge of CO emission defining the bottom rim of GSH 287+04--17. 
It is by far the largest molecular cloud associated with GSH 287+04--17 \citep[the associated portions of the GMC comprise $\sim60$\% of the total H$_2$ mass in the shell,][]{dawson08b,dawson08a} and also represents a substantial fraction of the gas in the vicinity of the Carina OB2 supershell. The cloud forms a contiguous structure in $l$--$b$--$v$ (spatio-velocity) space, with the bulk of the emission genuinely well-connected in the spatio-velocity domain -- i.e. the apparent connectedness is not the result of distinctly separated velocity components with marginally overlapping line wings. Its spatio-velocity structure is striking, and clearly shows that portions of the cloud are distinctly associated with one or both of the shells. This is illustrated in Figure \ref{gmc_lv}. 

\begin{figure}
\includegraphics[scale=0.57]{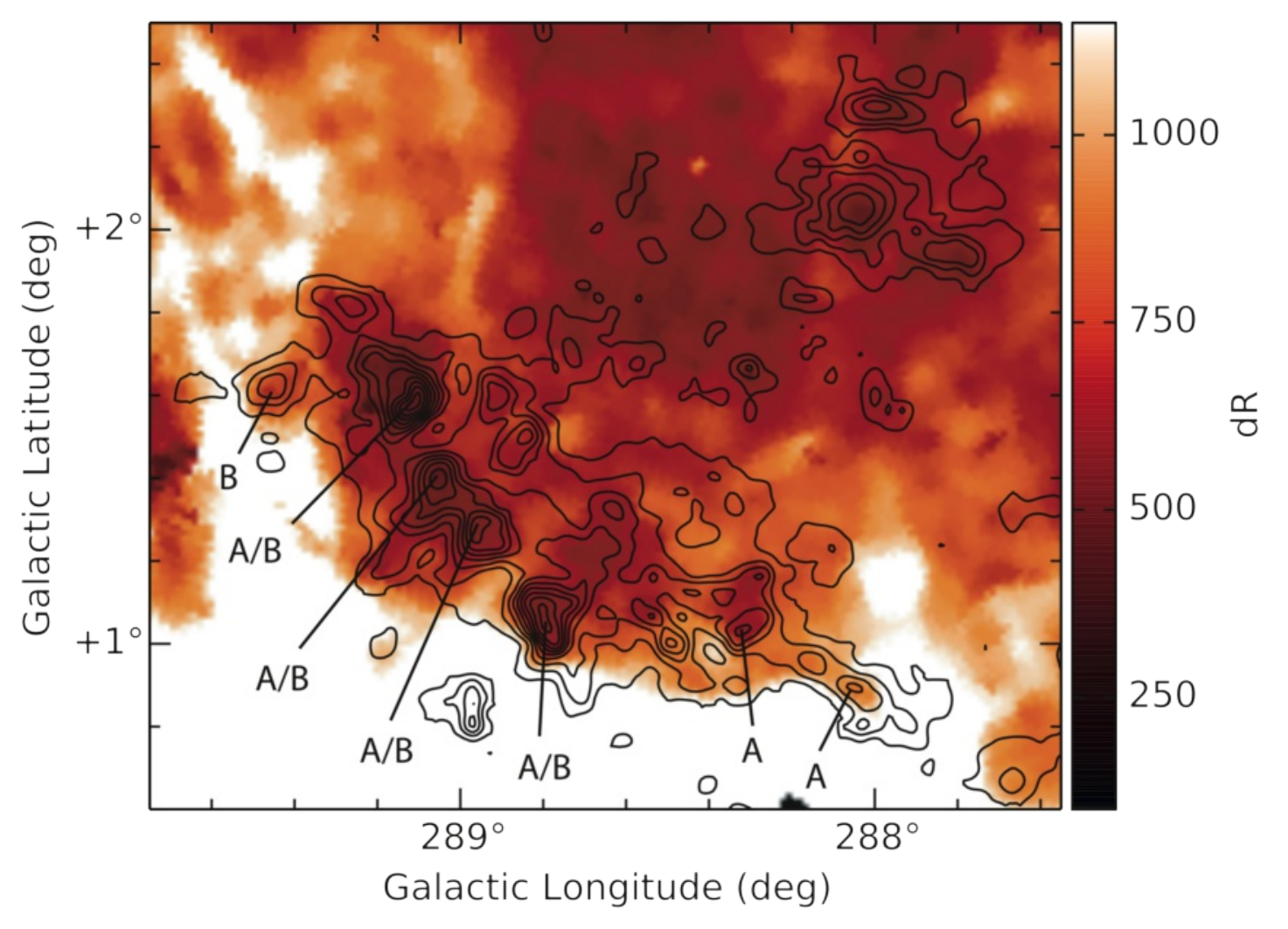}
\caption{$^{13}$CO(J=1--0) contours overlaid on an H$\alpha$ intensity map from the SHASSA survey \citep{gaustad01}. Contours are velocity-integrated intensity, integrated over all velocity channels where $^{12}$CO was detected at the $3\sigma$ level, drawn at intervals of 1.0 K km s$^{-1}$ and beginning at 0.5 K km s$^{-1}$. Note that the GMC is seen in absorption against the H$\alpha$. Emission peaks labelled `A' and `B' indicate those that are best interpreted as primarily associated with GSH 287+04--17 and the OB2 supershell, respectively. Peaks labelled `A/B' are either broad/complex profiles showing evidence of association with both objects, or show one component associated with each shell. The units of the image are deci-Rayleighs (dR)}
\label{gmc_shassa}
\end{figure}

Additional lines of evidence support the interpretation that the molecular gas genuinely occupies the same region of space and is not a chance superposition of unrelated components along the line of sight. The bulk of the $^{12}$CO(J=1--0) emission shows consistently low peak brightness temperatures of $\sim5$--6 K -- unusual for Galactic GMCs surveyed with the same instrument in similar regions of the Galactic Plane at similar distances \citep{mizuno04}, which typically show prominent subregions of brighter emission ($\sim10$--30 K), presumably associated with star formation activity. While not conclusive taken alone, this supports the idea that the emission is genuinely part of the same physical system, with similar properties. 
The cloud is also seen in absorption against H$\alpha$ emission lying at an estimated distance of 2.9 kpc \citep[][see also \citealt{dawson08a}]{georgelin00}. This absorption arises from the entirety of the GMC, including material associated with both shells, indicating that all of the molecular gas lies on the same side of the H{\sc ii} regions (see Figure \ref{gmc_shassa}). 
Finally, the alternative hypothesis requires that the location of this unusually large mass of molecular material exactly at the interface of the two shells 
be entirely coincidental. While this is not outside the realms of possibility, a causal relationship is strongly suggested, and bears investigation. 

\begin{table*}
\caption{Estimated properties of the two supershells}
\label{shellstable}
\begin{tabular}{@{}lcccccccccl}
\hline
Name & Size & D & $v_{\mathrm{lsr}}$ & $v_{\mathrm{exp}}$ & $M_{\mathrm{HI}}$ & $M_{\mathrm{H}_2}$ & $E_{\mathrm{kin}}$ & $E_{\mathrm{F}}^{\mathrm{~d}}$ & $\tau^{\mathrm{~e}}$ & Reference \\
 & (pc) & (kpc) & (km s$^{-1}$) & (km s$^{-1}$) & ($10^5~M_{\odot}$) & ($10^5~M_{\odot}$) & ($10^{50}$~erg) & ($10^{50}$~erg) & (Myr) \\
\hline
GSH 287+04--17 & $150\times230$ & $2.6\pm0.4$ & $-17$ & $\sim10$ & $7\pm3$ & $2.0\pm0.6$ &  $\sim10$ & $\sim50$ & $\sim10$ & \citet{dawson08a}\\
Carina OB2 & $80\times130$ & $2.9\pm0.9^{\mathrm{a}}$ & $-27$ & 22 & 1.1 $^{\mathrm{b}}$ & ... $^{\mathrm{c}}$ & $7.1$ $^{\mathrm{b}}$ & $\sim50$ & 4.1$^{\mathrm{b}}$ & \citet{rizzo98}\\
\hline
\end{tabular}\\
\begin{flushleft}
Notes:\\
a) Based on photometric distance estimates to the OB2 association \citep{garcia94,kaltcheva98,georgelin00,kaltcheva10}, with the uncertainty taken from the large distance spread found by \citet{kaltcheva98} and \citet{kaltcheva10}.\\
b) Derived assuming a distance of 3.1 kpc.\\
c) Molecular gas is associated but its mass not estimated in the literature. \\
d) Estimated formation energy \\
e) Defined as effective radius divided by expansion velocity.
\end{flushleft}
\end{table*}


\subsection{Observational and Physical Properties of the GMC}
\label{obsprop}

\begin{figure}
\begin{center}
\includegraphics[scale=0.8]{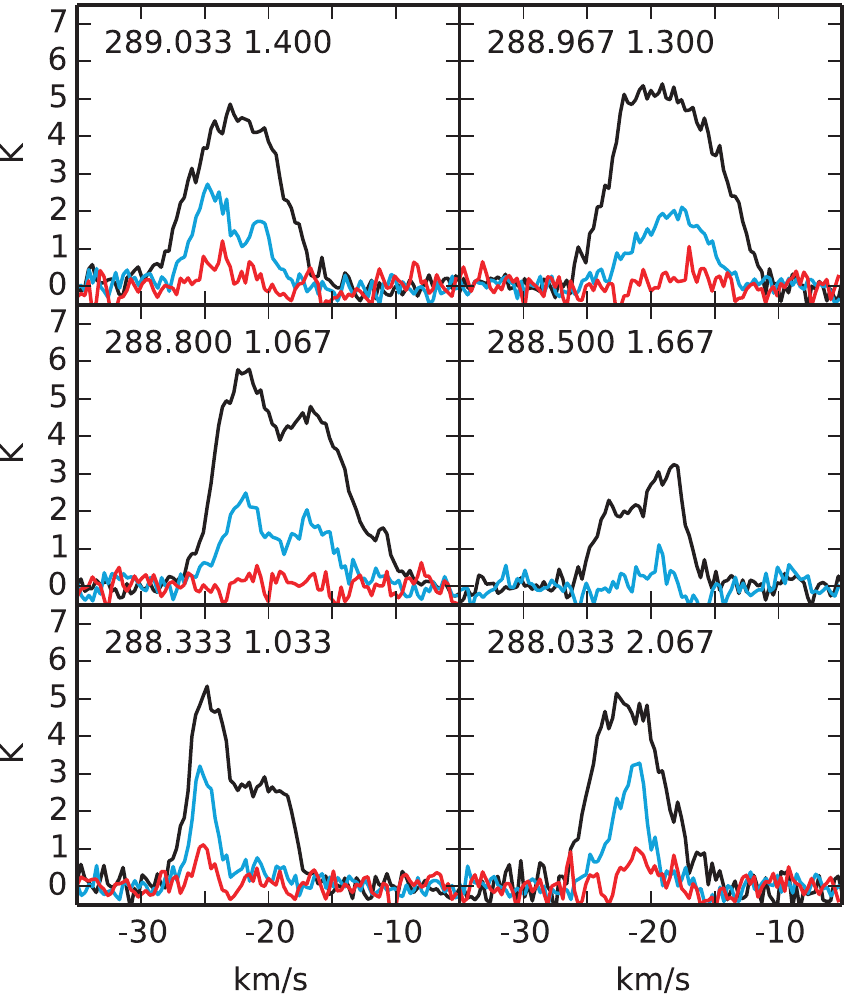}
\caption{NANTEN brightness temperature spectra for $^{12}$CO(J=1--0) (black), $^{13}$CO(J=1--0) (blue, scaled by a factor of 2) and C$^{18}$O(J=1--0) (red, scaled by a factor of 5) at selected integrated intensity peaks of the GMC. The numbers in the upper left-hand corners of each panel indicate the Galactic longitude and latitude at which each spectrum was taken.}
\label{gmc_specs}
\end{center}
\end{figure}

Emission from GMC 288.5+1.5 lies in the range $-33 < v_{\mathrm{LSR}} < -11$ km s$^{-1}$, with line profiles that are both broad and complex. Spectra at single spatial positions show evidence of broad ($\sigma_v\sim4$ km s$^{-1}$) and apparently single-peaked components, as well as examples of multiple blended velocity components (examples are given in Figure \ref{gmc_specs}). The intensity-weighted velocity dispersion for the entire cloud in $^{12}$CO(J=1--0) is $\sigma_v(\mathrm{cld})\approx3.7$ km s$^{-1}$, which, despite the apparently dynamically disrupted nature of the gas, is typical for Milky Way GMCs of this size \citep[e.g.][]{larson81,solomon87}.  

The GMC shows a bright ridge that extends for a projected length of $\sim90$ pc (assuming $D=2.6$ kpc) along the interface of the two supershells. This ridge is strongly detected in $^{13}$CO(J=1--0), indicating the presence of relatively dense and high column density gas. A $^{12}$CO envelope extends to the North-West, joining the ridge to a second concentration of $^{13}$CO-bright material projected within the boundary of GSH 287+04--17. The $^{12}$CO(J=1--0)/$^{13}$CO(J=1--0) peak brightness temperature ratio ranges from $\sim3$ at positions of peak $^{13}$CO intensity to typical values of $\sim10$--15 in the $^{13}$CO-poor zone, where these values are computed from mean spectra summed over different regions of the cloud. This latter value illustrates that 
a significant portion of the GMC is comprised of relatively diffuse molecular material \citep[see e.g.][]{polk88}. 

The C$^{18}$O line is below ($3\sigma$) detectability in all but a handful of individual spatio-velocity pixels (``voxels''); however, weak emission below the detection threshold is recovered when summing $^{13}$CO-detected voxels only, which indicates the presence of at least some dense ($n_{\mathrm{H}_2}\sim10^4$ cm$^{-3}$) gas. This denser material is seen in isolated clumps distributed along along the length of the bright ridge, as well as in $^{13}$CO peaks in the North-West structure.  

The H$_2$ mass traced in the $^{12}$CO, $^{13}$CO and C$^{18}$O lines is estimated to be $M_{\mathrm{H}_2}(^{12}\mathrm{CO})\sim1.7\times10^5~M_{\odot}$, $M_{\mathrm{H}_2}(^{13}\mathrm{CO})\sim3.5\times10^4~M_{\odot}$ and $M_{\mathrm{H}_2}(\mathrm{C}^{18}\mathrm{O})\sim0.8\pm0.4\times10^4~M_{\odot}$. The $^{12}$CO-based mass is estimated directly from the integrated intensity over the velocity range of the cloud, assuming a Galactic X-factor of $2.0\times10^{20}$ \citep{bolatto13}.  
For $^{13}$CO(J=1--0) the datacube is first masked to include only voxels detected at the $3\sigma$ level in the $^{12}$CO line, 
and $^{13}$CO column densities 
computed for each voxel from the standard LTE (local thermodynamic equilibrium) 
expressions \citep[e.g.][]{dawson11b}, with an assumed excitation temperature of $T_{\mathrm{ex}}=10$ K. The final conversion to $N_{\mathrm{H}_2}$ and the H$_2$ mass assumes an H$_2$-to-$^{13}$CO abundance ratio of $5\times10^5$ 
\citep{dickman78}. A similar method is employed for C$^{18}$O(J=1--0), with the cube masked to include only $^{13}$CO detections, and an assumed abundance ratio of $6\times10^6$ \citep{frerking82}. The large uncertainty quoted for the C$^{18}$O mass reflects the weakness of the emission relative to fluctuations in the spectral baselines for this line.

\subsection{Star Formation Activity in the GMC}
\label{starformation}

The weakness of the C$^{18}$O(J=1--0) emission and the low brightness temperatures in the $^{12}$CO line suggest that the level of star formation in the GMC is likely to be low; the former suggests that the dense gas fraction is relatively small, and the latter suggests an absence of strong heating sources within the cloud. For illustration, the properties of GMC 288.5+0.5 may be compared with those of the nearby very active star forming region $\eta$ Carina GMC \citep[also observed with NANTEN by][]{yonekura05} in which $^{12}$CO(J=1--0) line peak temperatures are $\gtrsim20$ K throughout much of the cloud and the mass traced in C$^{18}$O(J=1--0) comprises $\sim18\%$ of the $^{12}$CO mass ($44\%$ of the $^{13}$CO mass), compared to $\sim5\%$ (and $\sim23\%$) in the present case. The absence of visible (c.f. Figure \ref{gmc_shassa}) or radio (843 MHz, \citealt{mauch03}; 1.4 GHz, \citealt{haverkorn06}) H{\sc ii} regions, as well as a lack of maser tracers of massive star formation (6.7 GHz methanol, \citealt{green12}; 1667/1665 MHz OH, \citealt{caswell98}) support a picture in which the level of massive star formation activity is low.

\begin{figure}
\includegraphics[scale=0.65]{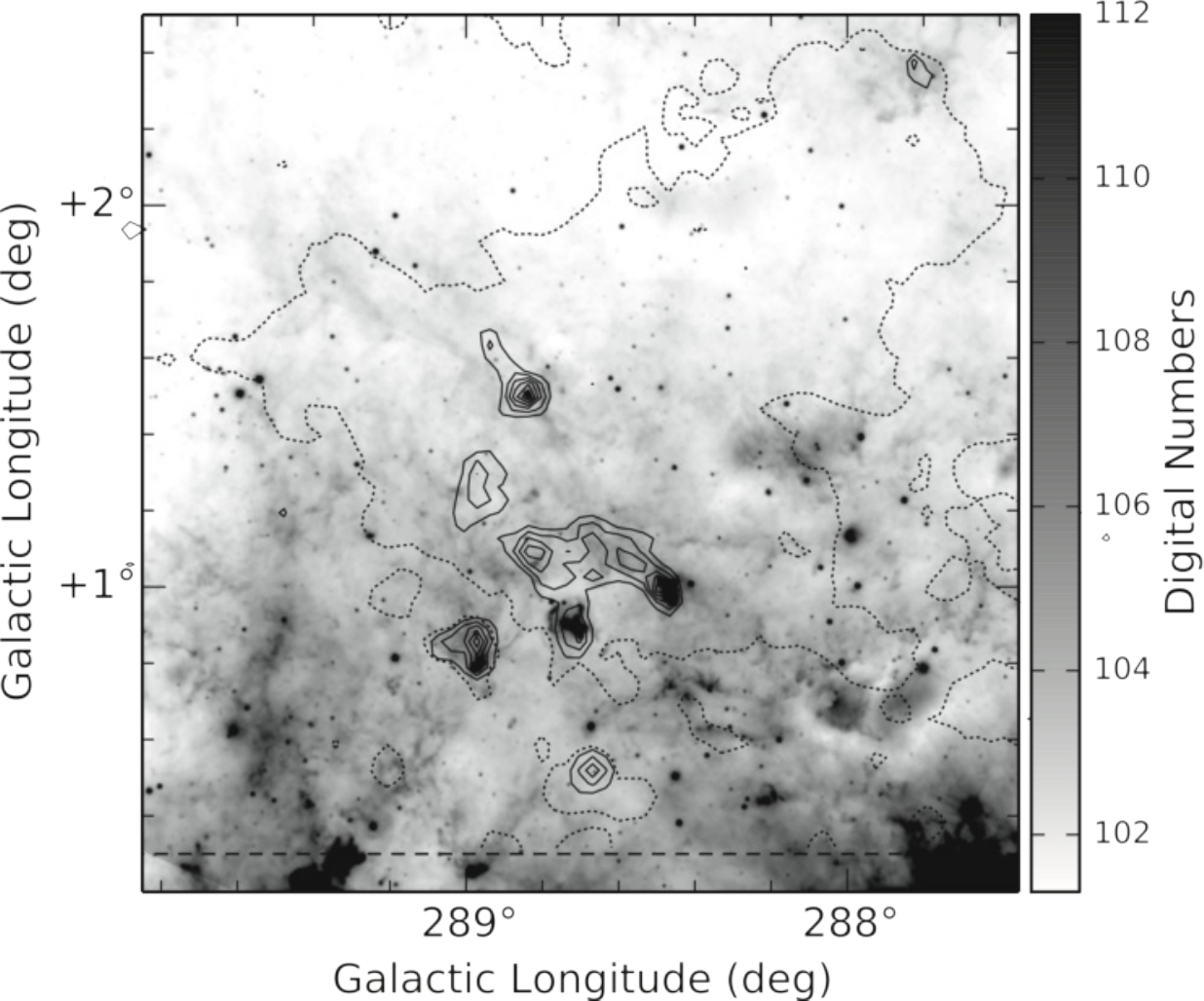}
\caption{WISE \citep{wright10} $22~\mu$m greyscale image of the region around the GMC, showing the warm dust indicative of star formation activity. The dotted contours show $^{12}$CO(J=1--0) emission integrated over the full velocity extent of the GMC ($-33<v_{\mathrm{LSR}}<-10$ km s$^{-1}$), at a level of 3.5 K km s$^{-1}$. Black contours are $^{13}$CO(J=1--0) emission limited to the velocity range in which the gas shows good spatial correlation with the dust emission ($-16.1<v_{\mathrm{LSR}}<-10.4$ km s$^{-1}$). These contours are drawn every 0.6 K km s$^{-1}$. It can be seen that the four compact $^{13}$CO clumps show a strong correlation with the only bright $22~\mu$m features in the region, but that the majority of the GMC shows little evidence for localised gas heating. The dashed line marks the extent of the observed region.}
\label{gmc_wise}
\end{figure}

Nevertheless, there are some signs of star formation in the GMC. Figure \ref{gmc_wise} shows $22~\mu$m data from the Wide-field Infrared Survey Explorer \citep[WISE;][]{wright10}, 
which is a tracer of warm dust heated by young stellar objects (YSOs). 
The majority of the projected area of the GMC shows little evidence for excess $22~\mu$m emission, consistent with minimal or low-level ongoing star formation. However, there are four bright, spatially resolved $22~\mu$m features, each of which has a bright $^{13}$CO molecular counterpart located between $-16$ and $-10$ km s$^{-1}$. It is notable that while these features are regarded part of the GMC complex (with the possible exception of the clump at $l\approx289.0^{\circ}$, $b\approx+0.8^{\circ}$, which is fully isolated in $l$-$b$-$v$ space), and are positioned consistently 
with a location on the shell rim, they are nevertheless somewhat distinct from the larger agglomeration of the main cloud, both in terms of their location in velocity and in terms of their properties: Each shows a single, narrow ($\sigma_v\sim1$ km s$^{-1}$) velocity component with peak brightness temperature of $\sim6$--8 K -- the highest values in the dataset. 
We therefore conclude that while (to the limits of this simplistic analysis) some star formation is clearly occurring, it is localised, 
with the bulk of the molecular gas apparently unaffected. 






\section{Numerical Simulations}
\label{model}

The formation of dense gas at the interface of two colliding superbubbles has been studied in 2-dimensional hydrodynamical simulations by \citet{ntormousi11}. Here we present an improved 3-dimensional implementation of the same models. 
These are not intended to specifically model GSH 287+04--17 and the Carina OB2 supershell, nor to recreate the sequence of events leading to the formation (or not) of the G288.5+1.5 giant molecular cloud. They do, however, provide a meaningful theoretical counterpoint against which to compare and contrast the observational results of this work. 

A full description of the models and their interpretation will be presented in Ntormousi et al. (in prep).

\subsection{Code and additional model implementation}

We model two colliding superbubbles in three dimensions using the Adaptive Mesh Refinement (AMR) hydrodynamics code RAMSES \citep{Teyssier02}, suitably adapted to simulate the feedback from young stellar populations.  The superbubbles are created by thermal and kinetic feedback from OB associations, which is approximated in the code by equally distributing the total thermal energy and mass output from 30 stars among a group of cells inside a spherical region of 5 pc radius.  More details about the wind implementation in RAMSES can be found in \cite{Fierlinger12}.

The feedback masses and energies have been calculated and provided to us by \citet{Voss09} as the average output from a typical galactic OB association, including stellar winds and supernovae.  Although the average UV radiative feedback is also available in the data, radiative effects are not simulated in our models due to lack of computing power.  However, we have performed simple one-dimensional calculations 
which show that, apart from the very first stages of the expansion, the radiation front lies behind the shock front for the duration of the models.

In order to create a two-phase medium we must simulate the cooling and heating processes in the ISM.
A very detailed description of these processes is given in \cite{Wolfire95} and we have used the rates from that work here in a tabulated form as a function of density and temperature for a gas of solar metal abundance.
Both the wind module and the cooling and heating function are the same as in \citet{ntormousi11}.

While molecular chemistry is not implemented in this version of the models, their sub-parsec resolution is sufficient to 
follow the evolution of the ISM into the  temperature and density regime where molecule formation would normally occur. Models implementing full chemical networks with non-LTE reactions have recently demonstrated that (a) molecules form quickly in a turbulent medium once sufficient densities are reached \citep{glover10}, and (b) the presence or absence of molecular cooling in fact has little effect on the ability of the ISM to cool and produce star-forming gas \citep{glover12}. We therefore feel justified in regarding ``dense gas'' ($n_\mathrm{H} \gtrsim 100$ cm$^{-3}$) as loosely equivalent to ``molecular gas'' for the purposes of interpreting the present simulations.

\subsection{Initial conditions} 

Instead of an idealized, homogeneous background we model the expansion of the supershells in a structured warm ISM.  Previous calculations \citep{ntormousi11} have shown that modeling supershell expansion in a turbulent rather than in a homogeneous environment produces more structured dense clumps, with morphologies and velocity dispersions much closer to observed clouds.

The initial turbulence is created using the \citet{MacLow_1999} recipe, which first introduces random (Gaussian) phases to four wavenumbers (k=1--4) in Fourier space and then takes an inverse Fourier transform to create the three components of the velocity.  This velocity field is applied to a box with homogeneous density and integrated in time long enough for the density-weighted power spectra of the turbulence to approach Kolmogorov behavior.

The simulation box has a physical size of 200 pc$^3$ and the resolution is uniform, equal to 512$^3$ grid points.  The feedback areas are placed on either side of the box.  The mean density of the warm medium is 1 cm$^{-3}$ and the temperature is set to the equilibrium of the cooling curve for this density, which is 8000 K.

\begin{figure*}
    \includegraphics[scale=0.5]{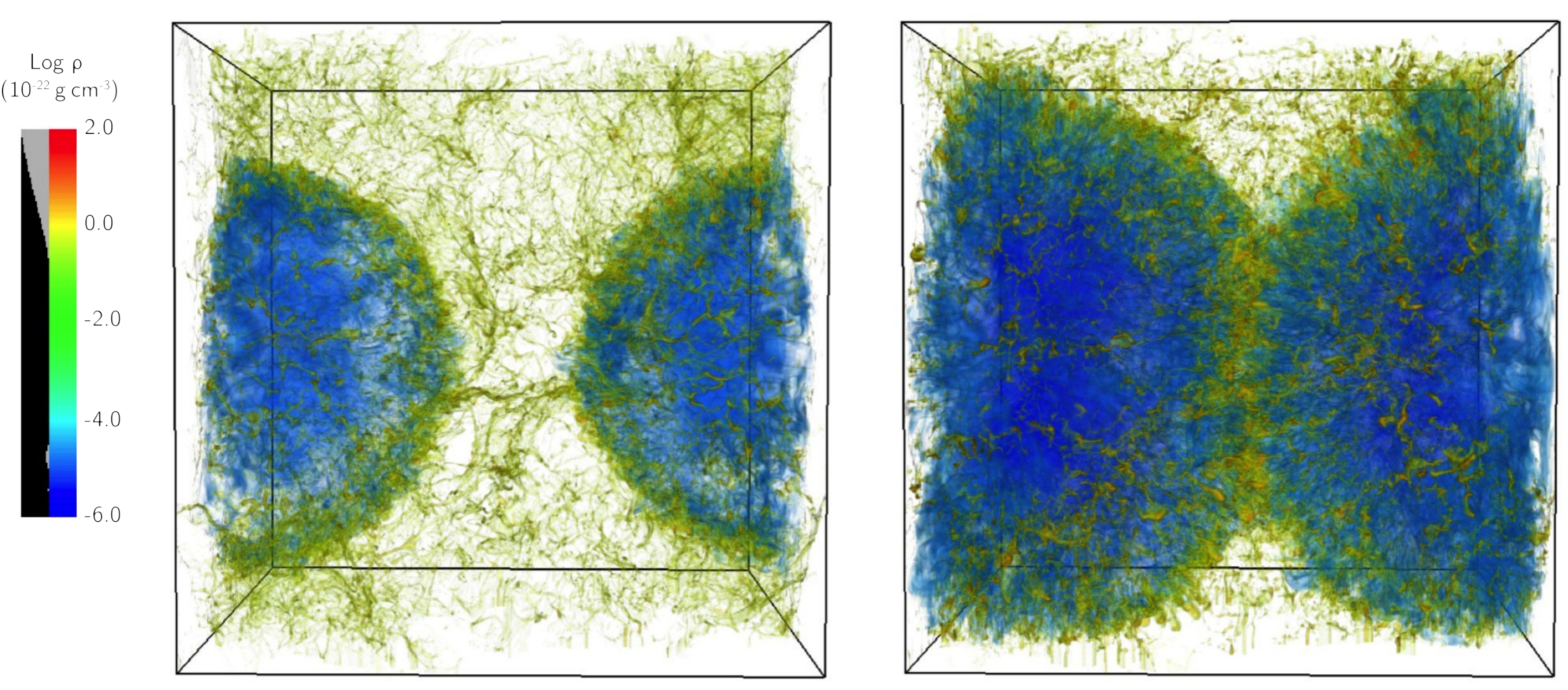} 
    \caption{Volume rendering of the logarithm of the density field in the 3D hydrodynamical simulations.  Left panel: snapshot before the shell collision, about 3.5 Myr after the start of feedback. Right panel: snapshot about 1 Myr after the start of the shell collision (5.5 Myr after the start of feedback).  The warm medium in this figure is made transparent to allow view of the dense clumps, as indicated by the grey/black curve along the colorbar.  The mass density units assumed here are 10$^{-22}$ g cm$^{-3}$ ($\approx60$ cm$^{-3}$) and the size of the box is the total 200 pc simulated.}
     \label{shells_volume_rendering}
\end{figure*}

\subsection{Model evolution}

Two models are presented here for comparison with the observations, one including self-gravity and one with hydrodynamics only.  
Within the timescales simulated by these models we actually do not expect gravity to have had time to act, apart from maybe on the densest structures.  
As an order-of-magnitude estimate, the free-fall time $t_{ff}=\left(\frac{3\pi}{32G\rho_0}\right)^{1/2}$ of the dense gas ($n_H\simeq100$ cm$^{-3}$) is about 5 Myr and only becomes
1 Myr for structures with $n_H\simeq1000$ cm$^{-3}$, not taking into account their internal velocities. The superbubbles evolve and collide in the middle of the computational volume 4.55 Myr after the feedback started at the edges of the box and we integrate the models for
about 2 Myr more.  Further evolution of the models, apart from being computationally very demanding, could also lead to contamination from the boundaries.
In fact, the evolution and the properties of the dense gas in the two models are very similar. 

In the same way as in the two-dimensional models in \citet{ntormousi11}, 
various fluid instabilities create rich structure on the surface of the dense shocks and cause them to fragment into small clumps.
Perhaps the most dynamical of these processes is the Vishniac instability, which is expected to happen on a dense spherical shock \citep{Vishniac_83}.
Very reminiscent of the Non-linear Thin Shell Instability (NTSI), described by \citet{Vishniac_94}, this instability focuses material on the tips of ripples
along the surface of the shell, in certain cases also triggering the growth of other fluid instabilities.  
In this particular environment, the shear inside the shock gives rise to small-scale Kelvin-Helmholtz eddies which contribute to the kinematics of the clumps, 
while the condensation at the tips of the ripples makes the gas thermally unstable, causing it to cool further until it reaches the second equilibrium point of the ISM cooling-heating curve 
(about 100 cm$^{-3}$ and $100$ K).  
At the same time, the deceleration of the shock causes Rayleigh-Taylor-like fingers to form on the inner surface of the supershells.  The resulting clumplets, exposed to the hot high-velocity winds in the interior of the superbubble
get gradually evaporated and acquire a head-tail structure. These instabilities are discussed further in the context of the observations in section \ref{instabilities}.

A three-dimensional view of the setup and the morphology of the clumps is seen in Figure \ref{shells_volume_rendering}, where two snapshots have been chosen:
one at about 3.5 Myr into the evolution of the model and another at about 5.5 Myr, when the two bubbles have already started colliding.   These two snapshots will serve as examples 
for the remainder of this Section.
The color-coding in these plots corresponds to the logarithm of the density in code units, which in this case is 10 $^{-22}$ gr/cm$^{-3}$.
It is clear that the shells become very dynamic and, although some dense structure is present in the surrounding medium due to local overdensities
in the turbulent warm ISM, the clumps on the surfaces of the shocks are denser and larger.  Although some agglomeration and merging of clumps certainly
happens on the shells as they sweep up these pre-existing overdensities, 
a comparison of these simulations to simulations of the same environment without the shocks produced almost no dense gas. We also note that the effect of self gravity in these models is indeed negligible, in accordance to the simple order-of-magnitude calculation above; there is no measurable effect on the dense gas distribution by turning on self gravity in the simulation.

\section{Discussion}

\subsection{Origin of the Molecular Gas}
\label{origins}

GSH 287+04--17 contains a considerable amount of associated molecular gas -- more than would be expected from comparisons with the non-disturbed ISM in its local vicinity \citep{dawson11a}. It has already been argued based on these results that as much as half of the molecular mass associated with the shell (of which the associated portions of the GMC comprise $\sim60\%$) may have been formed from the atomic medium by the accumulation of material in the shell walls. The GMC is notable as both by far the largest and the most massive concentration of molecular gas associated with GSH 287+04--17. Other associated clouds, even those at similar Galactic latitudes, have typical masses of $M_{\mathrm{H}_2}(^{12}\mathrm{CO})\lesssim10^4~M_{\odot}$ \citep{dawson08b} -- more than an order of magnitude smaller.

\begin{figure*}
\begin{center}
   \includegraphics[scale=0.87]{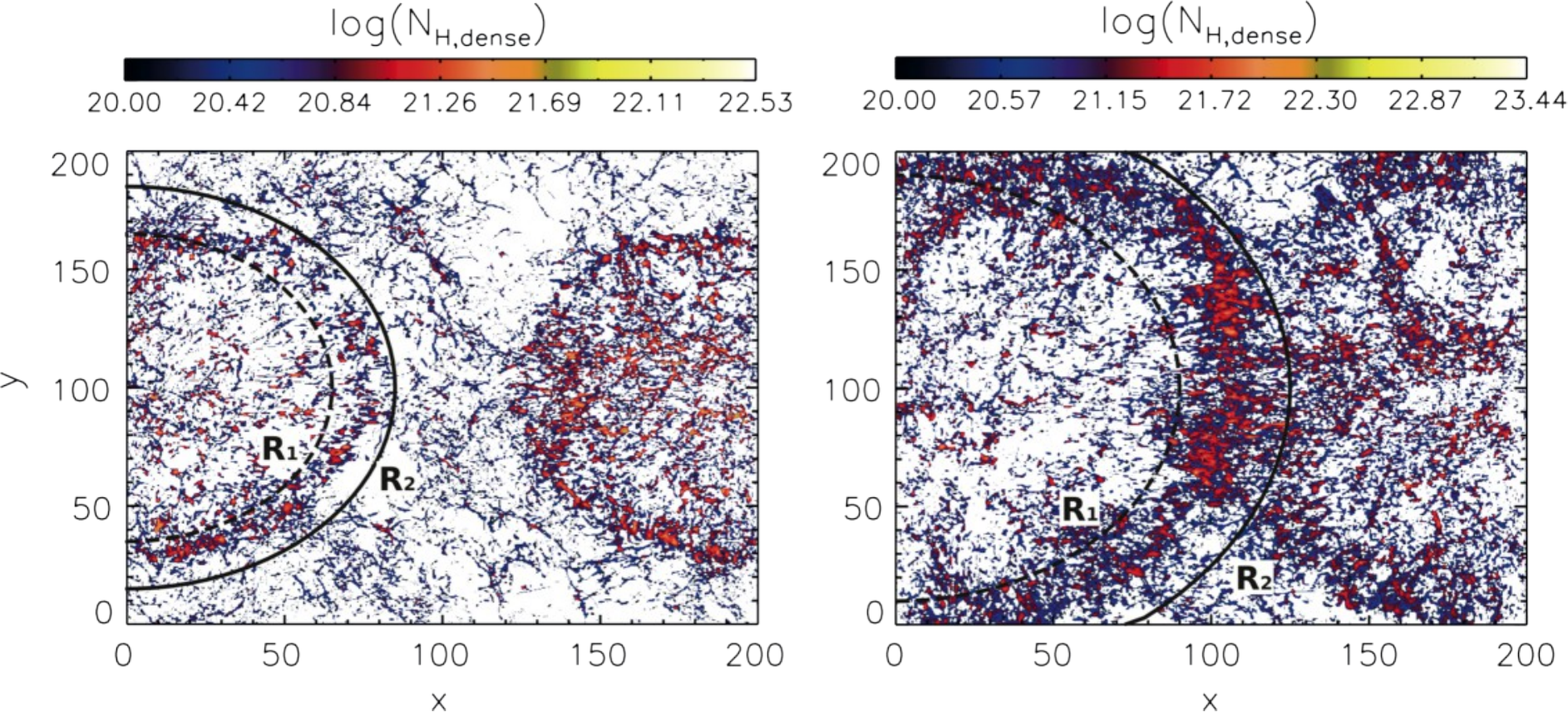}
    \caption{
    Column density of the dense gas only ($n_H>100$ cm$^{-3}$) integrated along the z-direction in the pure hydrodynamic runs of the simulations. The left panel shows a 3.5 Myr snapshot and the right panel shows a 5.5 Myr snapshot (after the shells have collided). The axes are in parsecs. 
    The R$_1$ and R$_2$ labels refer to the radii ranges plotted in figure \ref{1Dmass_plots}.
    These figures are for the pure hydrodynamic run, but there is no significant difference to the self-gravitating run.}
     \label{column_densities}
     \end{center}
\end{figure*}

\begin{figure}
    \includegraphics[scale=0.31]{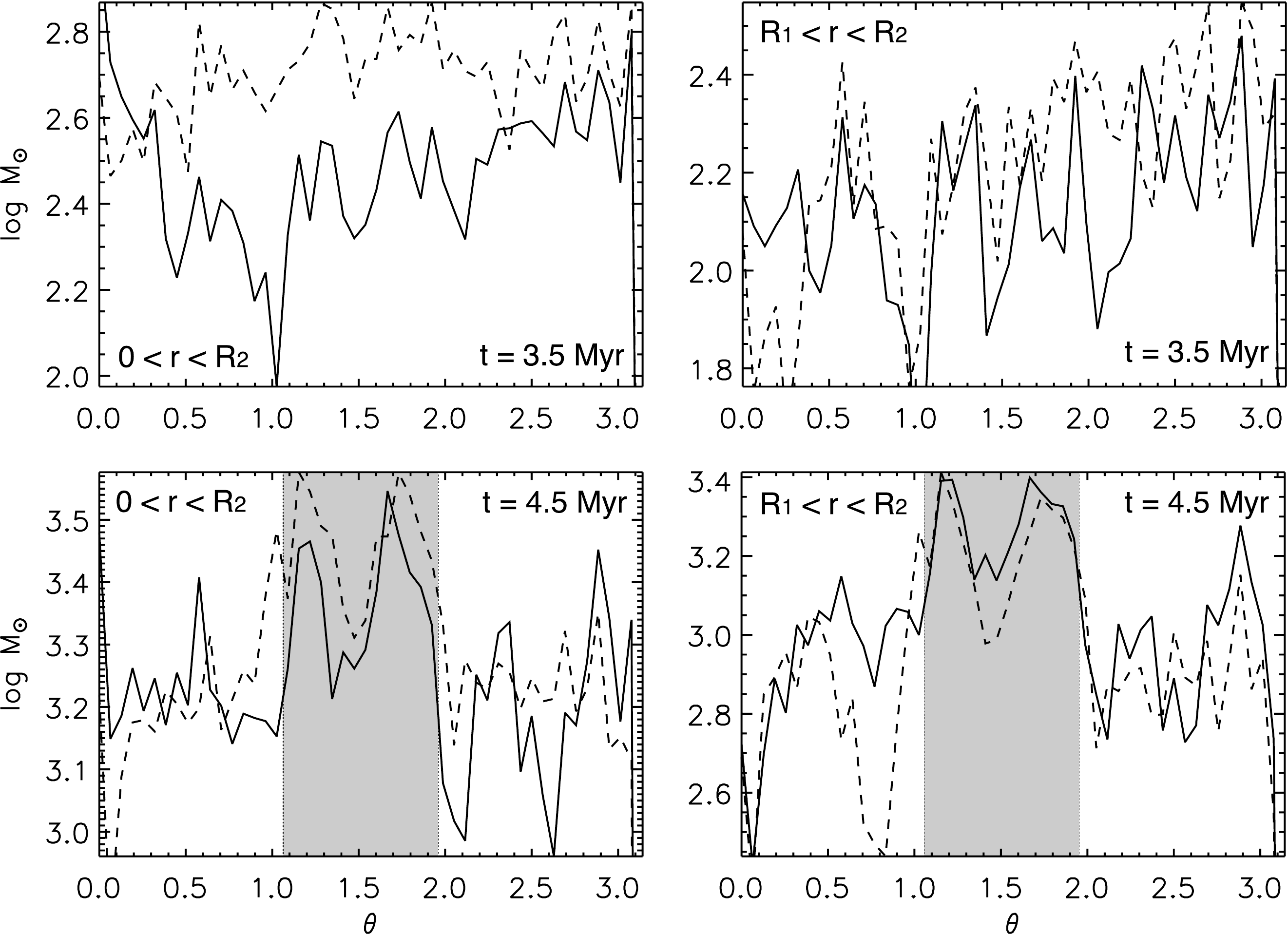}
    \caption{Dense gas mass ($n_\mathrm{H} > 100$ cm$^{-3}$) in azimuthal bins ($\theta$ to $\theta+d\theta$) as a function of the angle $\theta$ around the centre of the simulated bubbles, as derived directly from the column density maps in Figure \ref{column_densities}. Solid lines show results for the left-hand bubble and dashed lines for the right-hand bubble. The top panels show plots for the early stage snapshot (3.5 Myr) and the bottom panel for the post-collision snapshot (5.5 Myr). The left-hand panels show the mass contained in ``pie slices'' of radius 0 to R$_{2}$ (as indicated in Figure \ref{column_densities}), while on the right only the mass in a ring from R$_{1}$ to R$_{2}$ is calculated. Grey shaded areas mark the approximate location of the collision zone. These plots are for the pure hydrodynamic run only, since the results from the self-gravitating run are almost identical.} 
     \label{1Dmass_plots}
\end{figure}

In the numerical models, the shell collision does not appear to create either (a) much larger or (b) denser structures than those present on the shells outside the collision zone. 
Figure \ref{column_densities} shows column density maps of the dense gas ($n_H>100$ cm$^{-3}$) before and after the shell collision, and Figure \ref{1Dmass_plots} shows this mass as a function of angle $\theta$ around the centers of the two bubbles. 
While there is undoubtedly a concentration of dense material at the collision zone, the mass increase is typically only around a factor of two, and can be simply interpreted as the addition of the material already present on the surfaces of the the two shells -- 
i.e. there is no evidence for \textit{enhanced} dense gas formation due to the collision. 

In fact, the model clumps appear to be disrupted by the violent merging of the interior gas flows. While the expansion velocities of the model shells prior to the collision are a moderate $\sim10$--15 km s$^{-1}$ (close to the observational values of 10--20 km s$^{-1}$, and typical of the general population of similarly sized Galactic supershells; \citealt{heiles79,mcclure02}), the interior gas flows are hot, fast and disruptive. The innermost regions of the superbubbles are filled with $\sim10^7$ K gas, with velocities over 1000 km s$^{-1}$, and any clumps exposed to this hot phase will be evaporated. However, even the far more distant clumps within the collision interface are not safe -- they 
are compressed by cooler, denser ($\sim1$--10 cm$^{-3}$) flows of 20--40 km s$^{-1}$ from each side, and the resulting shear is enough to effectively strip them apart. 
This may indicate an overestimation of the effect of the feedback due to our inability 
to model the escape of hot, interior gas through pre-existing low-density regions (``chimney flows'') out of the plane of the Galaxy, a situation frequently encountered, both in observations and in global numerical models of disks with stellar feedback \citep[e.g.][]{avillez01,mcclure06,henley10,hill12}. Indeed, GSH 287+04--17 is known to be a Galactic Chimney system \citep{dawson08a}. Similarly, we are also unable (due to lack of resolution) to model either clumpy SN ejecta or individual wind collisions within the OB association -- both processes which could absorb some of the feedback energy. 

The lack of magnetic fields in our models is also likely to be relevant. Pure hydrodynamical models tend to find that dense gas forms far more efficiently 
than in the full MHD case \citep[see e.g.][]{inoue08,heitsch09,inoue09,vazquez11}. In our models 
dense gas production proceeds efficiently in the shell walls without the need for the additional compression and influx of material provided by the collision. However, if magnetic fields were present, we might expect dense gas formation to be inhibited, 
with the collision offering a means of overcoming the additional magnetic support, and hence potentially enhancing dense gas formation. 
These issues will be investigated in future MHD models (Ntormousi et al. in prep.). 

Nevertheless, it is clear from the observations 
that the collision of GSH 287+04--17 and the Carina OB2 supershell cannot have produced the GMC entirely from a canonical ``ambient atomic medium'' ($n\approx1$ cm$^{-3}$). 
Approximating the bright ridge of the cloud as a disk of diameter $\sim90$ pc sandwiched between the two shells, taking the shell centers as $l=290.1^{\circ}$, $b=+0.2^{\circ}$ and $l=287.5^{\circ}$, $b=+3.0^{\circ}$, and assuming a main ridge mass of $M_{\mathrm{H}_2}=1.2\times10^5~M_{\odot}$, we derive a mean initial number density for the pre-shell medium of $\langle n_\mathrm{H} \rangle \sim10$ cm$^{-3}$, implying that some pre-existing dense material was 
present prior to the formation of the GMC. 

This is not unexpected, particularly given the assumed location of the shells within a spiral arm. The Carina OB2 supershell in particular has pushed up through the Galactic midplane en-route to the collision zone, likely encountering some cool and dense material on its way. 
The nature of this material is unclear, however. At one extreme we may imagine a mixture of classical warm neutral medium (WNM, $n\sim0.5$ cm$^{-3}$) and cold neutral medium (CNM, $n\sim50$ cm$^{-3}$), which is driven entirely molecular by the action of the shells. This scenario would require the two phases to have relative volume filling factors of 5:1, which is higher than usually assumed for the Galactic ISM as a whole. 

At the other extreme, the GMC may have existed entirely in its present dense, molecular form since before its interaction with the two shells, with the shell material only contributing very minimally to its mass. While this scenario cannot be ruled out, we regard it as unlikely for the following reasons. 1. The GMC is located exactly at the collision zone, and is an order of magnitude more massive than other clouds in the vicinity. A pre-existing cloud -- even one that has been picked up and moved -- would be no more likely to fall between the shells than elsewhere. 
2. Evidence already exists for enhanced molecular gas formation in GSH 287+04--17 as a whole, based on an overabundance of CO within the shell region \citep{dawson11a}, and the GMC contributes to this overabundance. 
3. A number of recent studies suggest that molecular clouds form and evolve rapidly, with the onset of star formation happening within a few Myr after the formation of molecular gas \citep[e.g.][]{elmegreen00,elmegreen07,hartmann01,hartmann12,vazquez10}. If this picture is correct, then the fact that the GMC shows no significant star formation activity over most of its volume suggests that much of the gas is young and newly-formed. This is consistent with the age estimates of the two shells, which place the collision no more than a few Myr in the past. 


We favor a scenario in which GMC was seeded by pre-existing denser material, but was grown through the compression of this material at the shell interface. 
In a realistic ISM, this seeding material includes both CNM and some amount of molecular gas. 
Pre-existing molecular gas may include disrupted material from an original parent cloud, smaller structures encountered in the shell expansion, and existing material at the current location of the GMC. It is important to note that this gas need not remain molecular for the full duration of its interaction with the shells; indeed encounters between pre-existing clouds and supershells may well be disruptive \citep[see][]{dawson11a,dawson11b}. The role of the shells is then to bring this material together, enhancing the molecular gas fraction in the collision zone. 

This scenario is very consistent with the picture of molecular cloud formation recently outlined by \citet{inutsuka14}, in which the formation of GMCs requires multiple compressive events, often involves the assembly of smaller pre-existing dense structures, and happens commonly in the overlapping regions of superbubbles, where such multiple compressive events can occur. The typical formation timescales of $\sim10$ Myr suggested by that work are not inconsistent with the relatively recent collision of the observed shells, which in the \citet{inutsuka14} picture would only represent the final stage of a longer GMC assembly process.

\subsection{Gravitational Stability and Pressure Confinement}
\label{gravity}

A simple exploration of the gravitational stability of the GMC may be made by computing a `virial mass' as:
\begin{equation}
M_{vir}/M_{\odot} = 1160~R~[\sigma_v(\mathrm{cld})]^2 
\end{equation}
This gives the mass that would be required for self-gravity to just balance internal motions for a uniform spherical cloud of radius $R$ pc, a velocity dispersion of $\sigma_v(\mathrm{cld})$ km s$^{-1}$ 
in the absence of magnetic fields or external pressure. The $^{12}$CO and $^{13}$CO virial masses for the GMC have already been computed by \citet{dawson08b}. They find $M_{\mathrm{vir}}(^{12}\mathrm{CO})=4.7\times10^5~M_{\odot}$, which is revised to $5.6\times10^5~M_{\odot}$ based on our present definitions of the extent of the GMC. Here we have assumed $R=\sqrt{A/\pi}\approx35$ pc where $A$ is the projected area of the cloud in $^{12}$CO(J=1--0) emission, and $\sigma_v(\mathrm{cld})=3.7$ km s$^{-1}$ (see section \ref{obsprop}). This is significantly larger than the luminosity-based mass estimate of $2.3\times10^5~M_{\odot}$ (in which we have now included a factor of 1.35 to account for the presence of helium and heavier elements). The situation in $^{13}$CO is even more extreme. 
The virial mass for the largest discrete region of $^{13}$CO(J=1--0) emission in the GMC -- the bright ridge \citep[cloud 109r in][]{dawson08b} -- is $3.1\times10^5~M_{\odot}$. 
The luminosity-based mass for this feature is only $3.7\times10^4~M_{\odot}$ 
-- almost an order of magnitude smaller. The same situation holds for all discrete $^{13}$CO `clouds' catalogued within the GMC, none of which are virialized under this formulation \citep{dawson08b}. 

The GMC is therefore not globally self-gravitating under the standard virial treatment, a fact which was remarked upon when the object was first catalogued \citep{dawson08b}. This is unusual for such a massive GMC, the majority of which are found to be virialized under this formulation \citep[e.g.][]{solomon87,maloney90,heyer01}. While this simplified treatment is a crude tool for probing the true gravitational state of a molecular cloud, the fact remains that the balance of internal motions to luminosity-derived mass is large compared to other Galactic GMCs.

We may explore whether 
the thermal pressure of a hot interior medium (if one still exists), is sufficient to provide confining pressure for the cloud. With the inclusion of a surface pressure term in the virial theorem, we obtain 
\begin{equation}
P_S=\frac{1}{4\pi R^3}\left(3M[\sigma_v(\mathrm{cld})]^2 - \frac{3}{5}\frac{GM^2}{R}\right)
\end{equation}
\noindent as the surface pressure just required to confine a spherical cloud, which leads to the condition $P_S\gtrsim7\times10^{-12}$ g cm$^{-1}$ s$^{-2}$ for the assumed properties of the GMC. In classical models of bubble evolution \citep[e.g.][]{Weaver_77}, the expansion of the system is driven by overpressurization of shock-heated, hot interior gas. The model shells provide us with a convenient estimate of the properties of this interior gas at the time of the collision: $n\sim10^{-4}$ cm$^{-3}$ and $T\sim10^7$ K, leading to $P_S=nkT\sim1.4\times10^{-13}$ g cm$^{-1}$ s$^{-2}$. This is insufficient to confine the GMC. 
Whether the observed systems are indeed well described as bubbles of hot gas is unclear, however. In the case of GSH 287+04--17, while there is possible evidence of soft X-rays from the cavity \citep{fukui99}, the shell is undergoing chimney breakout, which may have resulted in the venting of hot material into the Halo \citep{dawson08a}. The Carina OB2 supershell is perhaps more likely to still contain hot ionized medium, since the powering cluster is known and still contains several or more O-type stars \citep{garcia94,kaltcheva98}.  

An alternative source of external pressure is ram pressure associated with the collision. In the model shells, warm gas in the collision zone typically has densities of 1--10 cm$^{-3}$ and velocities of 20--40 km s$^{-1}$ from each side, though the velocity field around the dense clumps is complex. Taking these numbers at face value results in ram pressure estimates ($\rho v^2$) of between $\sim7\times10^{-12}$ and $3\times10^{-10}$ g cm$^{-1}$ s$^{-2}$ -- more than sufficient to confine the GMC. 



One implication of these results is that it is possible to form a giant molecular cloud without the need for global self-gravity, with ram pressure (in this case) providing the external force needed to confine the gas. A scenario in which GMCs are in general not  gravitationally bound has been suggested as a means of explaining the low star formation efficiencies observed throughout the local universe \citep[e.g.][]{clark05, bonnell11, dobbs11b}. While the general applicability of this theory is unclear, it is interesting to note that G288.5+1.5 appears to present a convincing example of an externally-confined, non self-gravitating GMC. In this particular case, we might expect the bulk of the cloud to begin to disperse again once the supershell energy sources have switched off and external ram pressure to the region has ceased.




\begin{figure*}
\includegraphics[scale=0.47]{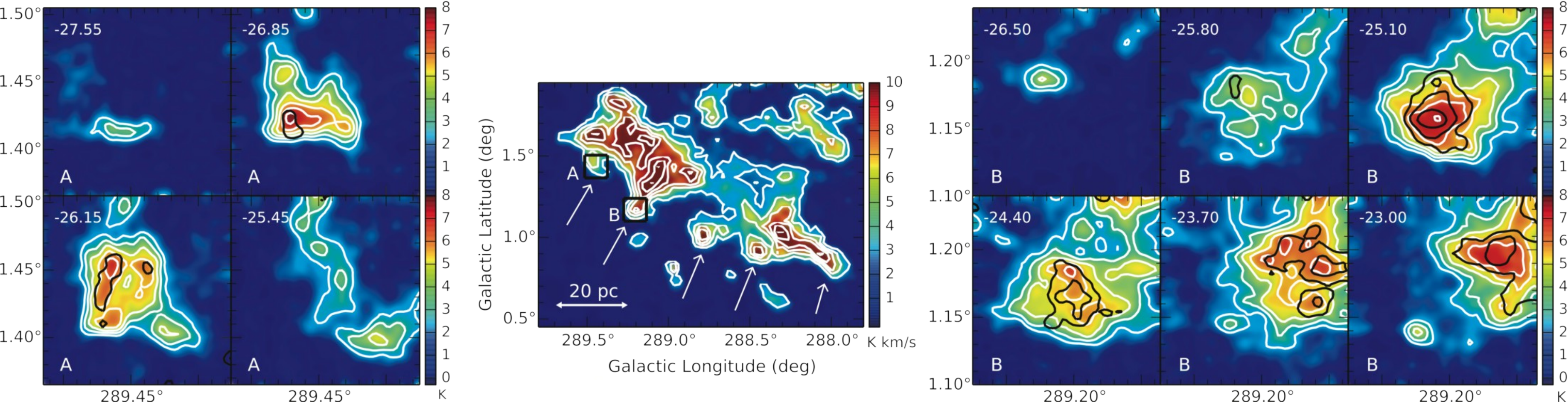}
\caption{Close-up of periodic drip-like structures along the GMC ridge. The central panel shows NANTEN $^{12}$CO(J=1--0) data integrated over the velocity range $-26.3 < v_{\mathrm{LSR}} < -23.0$ km s$^{-1}$. Both the color and the white contours show the same data, with the contour levels set at intervals of 2.5 K km s$^{-1}$. White arrows mark the locations of drip-like features spaced at approximately equal projected distances along the bottom ridge of the cloud. The black boxes labelled `A' and `B' mark areas observed at higher resolution with the Mopra telescope. The right and left panels show velocity channel maps of this higher resolution data. The colored images and white contours are $^{12}$CO(J=1--0), and black contours are $^{13}$CO(J=1--0). Contour levels begin at 2.5 K and are incremented every 1.0 K for $^{12}$CO, and begin at 0.8 K incremented every 0.5 K for $^{13}$CO. Numbers in the top left corner of each sub-panel indicate the central velocity of the displayed channel.}
\label{drips}
\end{figure*}

\subsection{Instabilities Along the Cloud Ridge}
\label{instabilities}


The GMC ridge shows periodic, `drip-like' formations of molecular gas spaced regularly along its bottom edge, where it meets the Carina OB2 supershell. These have a projected separation of $\sim20$ pc at the assumed distance of $2.6$ kpc, and apparent diameters of $\sim4$--5 pc. Figure \ref{drips} shows the NANTEN $^{12}$CO data integrated over the velocity range where these features are most prominent, as well as high-resolution ($\sim0.6$ pc) Mopra followup observations of two of these drips in $^{12}$CO(J=1--0) and $^{13}$CO(J=1--0). Both show evidence for dense $^{13}$CO heads, with thin ($\lesssim1$ pc) $^{12}$CO filaments joining them to the main body of the cloud. Emission along the bottom edges of the drips is sharp-edged, suggesting possible compression.

These drips, 
spaced almost equal projected distances from each other along the length of the cloud, 
are reminiscent of the fluid dynamical instability features that are seen throughout the dense gas in the numerical models.  It is therefore tempting to attempt to relate them to the corresponding growth rates of such instabilities in this environment, and try to interpret their presence in this position of the cloud. We note that discussion of gravitational fragmentation processes 
is not well justified here, 
since the large internal velocity dispersion of the cloud compared to its luminosity-based mass implies that the cloud is far from globally self-gravitating (see section \ref{gravity}). 

Assuming that both supershells still contain active energy sources, the GMC is being compressed between two high-Mach flows, which is a paradigmatic example of the non-linear thin shell instability \citep[NTSI,][]{Vishniac_94}.
This instability occurs in a slab supported internally by thermal pressure but bounded by ram pressure. It is 
the different nature of these pressures, the thermal being isotropic and the ram pressure 
being perpendicular to the flow, which creates an instability every time there is a ripple along the flows.  
A shear is created at the flow interaction region as material within the slab tends to be move towards the tips of the ripple. 

The growth of this instability in the classical case of two isothermal, non-turbulent flows may be approximated by
\begin{equation}
   \sigma_\mathrm{NTSI} \sim \left( \frac{\eta}{\lambda} \right)^{3/2} c_s~,
   \label{ntsi}
\end{equation}
where $\sigma_\mathrm{NTSI}$ is the growth speed, $\eta$ the amplitude of the perturbation and $\lambda$ is the perturbation wavelength \citep[using the formulation of Vishniac's equations given in][]{McLeod_Whitworth_2013}. Assuming $\lambda\approx20$ pc, and taking $\eta$ as the current perpendicular distance of the drips from the main body of the cloud ($\eta\approx5$ pc), we obtain growth speeds of 0.025 pc Myr$^{-1}$ for a 10 K molecular gas, and 1.2 pc Myr$^{-1}$ for an 8000 K warm atomic gas, corresponding to growth times of $\sim200$ and $\sim4$ Myr, respectively.
     
Of course, neither of these idealized scenarios is a realistic representation of the situation occurring at the shell interaction zone. Nevertheless, it is interesting to note that the growth of the NTSI in inflowing warm gas, possibly accompanied by a phase transition, is consistent with the simple calculation above. In this scenario, 
the drip-like structures would have been formed simultaneously with the cloud out of the instability of the interaction layer. However, this would require two warm flows with practically no pre-existing dense gas involved. As noted already, both the evolution of the numerical models and the initial density requirements for the formation of the GMC make this scenario highly implausible. Furthermore, as discussed by \citet{McLeod_Whitworth_2013}, supersonic turbulence in the incoming flows (quite plausible for a cooler gas) can act to effectively suppress the NTSI, introducing yet more complications to the simple picture outlined above. 


We may also consider the Rayleigh-Taylor (RT) instability, which occurs when a heavier fluid is accelerated by a lighter one. The edge of the cloud which hosts the drip-like formations faces a decelerating dilute flow, a configuration which is RT unstable. 
The RT growth rate is then expressed as 
\begin{equation}
\sigma_\mathrm{RT}=\sqrt{2\pi Aa/\lambda}~,
\end{equation}
where $A=(\rho_\mathrm{dense}-\rho_\mathrm{dilute})/(\rho_\mathrm{dense}+\rho_\mathrm{dilute})$, $\lambda$ is the wavelength of the perturbation and $a$ the acceleration. Approximating the main ridge of the cloud as a disk of mass $M=1.6\times10^5~M_{\odot}$ and a radius $R_\mathrm{cld}=90$ pc, sandwiched between the shells, and assuming again the model flow speeds and densities for the warm gas at the interaction zone (20--40 km s$^{-1}$ and 1--10 cm$^{-3}$), we may use the ram pressure computed in section \ref{gravity} to obtain an estimate of the acceleration of the dense gas. 
Taking $a=P_\mathrm{ram}\:\!\pi R_\mathrm{cld}^2/0.5M$ This is found to be between $\sim3\times10^{-9}$ and $\sim1\times10^{-7}$ cm s$^{-1}$, where we have assumed that half of the GMC mass is accelerated by the flows impacting its bottom side. Taking $A\approx1$, this gives growth times ($1/\sigma$) of $\sim0.3$--1.9 Myr. These timescales are very plausible, particularly given that the growth of RT features will likely have commenced before the onset of the collision itself. 


Indeed, as seen in the numerical models, it is likely that various fluid dynamical instabilities began developing at earlier epochs in the evolution of the two shells and grew to seed the structures now seen in the GMC. 
In this context is is interesting to note that some evidence of similar instability structures is seen throughout the walls of GSH 287+04--17, which have been observed at parsec resolution in both H{\sc i} and CO by \citet{dawson11a}. These authors find that the H{\sc i} shell walls show drip-like structures, the largest of which occur at longer wavelengths than the GMC drips, and many of which are tipped with denser molecular material. Similar long-wavelength drip-like features and ``scalloped'' structure have also been noted in the walls of another galactic shell, GSH 277+00+36 \citep{mcclure03}.  

One question, however, is why the $\lambda=20$ pc mode should be favored over smaller wavelengths with faster growth rates. A possible explanation is the magnetic field in the gas, which provides an effective surface tension that suppresses shorter wavelength modes. In the magnetized RT instability, the fastest growing wavelength is given by $\lambda_0=8\pi v_a / a$, where $v_a$ is the Alfven speed. For the range of accelerations computed above, this expression yields Alfven speeds of 1--5 km s$^{-1}$ for $\lambda_0=20$ pc, which are very reasonable for the cold ISM. The corresponding magnetized growth time is given by $2\sqrt{2}\:\!v_a/a$, which produces growth times of 0.4--2.5 Myr -- still very plausible compared to the shell lifetimes. One caveat is that these equations only hold when the magnetic field runs parallel to the interface between the two fluids; for the opposite direction the $B$ field has no effect in this classical formulation. However, 3D simulations by \citet{stone07} show that the field still ends up setting the preferred wavelength at later times, so the orientation of the field may not be a critical parameter.  



\section{Summary and Conclusions}
\label{conclusions}

Giant molecular clouds are often envisioned as forming at the stagnation points of large-scale ISM flows. Among proposed drivers of such flows is the stellar feedback from massive clusters, which drives repeated shock waves into the surrounding ISM, accumulating it into cold, dense shells. Recent work \citep{inutsuka14} has placed emphasis on the importance of multiple generations of shock compression, which are needed to provide additional material to a growing cloud, and to overcome the magnetic support that can prevent the transition from atomic to molecular gas. In this way, the interfaces of colliding supershells may be a particularly fertile ground for GMC formation. 

We have performed a detailed observational study of G288.5+1.5 -- a massive ($M_{\mathrm{H}_2}\sim1.7\times10^5~M_{\odot}$) GMC sandwiched between two Galactic supershells. This cloud may be the strongest candidate yet known for a GMC forming at the stagnation point of two feedback-driven flows; it shows robust evidence for genuine physical association with both objects, and several lines of evidence suggest that the molecular gas was assembled in its present location by the action of the shells. We have also combined this observational work with new 3D hydrodynamical simulations of superbubbles colliding in a turbulent medium, providing theoretical context that aids in the interpretation and analysis of our results. These models will be the focus of a dedicated upcoming paper (Ntormousi et al. in prep).

The present mass of the cloud, and the geometry of the two supershells, point to a scenario in which the GMC was formed in the collision zone from a combination of warm atomic gas together with some pre-existing denser material; though the nature of this material (cold atomic gas or smaller molecular clouds) is unknown. This is highly consistent with the picture of \citet{inutsuka14}, which stresses the importance of pre-existing dense gas in the formation of molecular clouds, envisioning a scenario in which early episodes of feedback form primarily cold H{\sc i}, which is later gathered into molecular clouds by subsequent generations of overlapping or colliding bubbles. The typical formation timescales of $\sim10$ Myr suggested by that work are not inconsistent with the relatively recent collision of the observed shells, which in the \citet{inutsuka14} picture would only represent the final stage of a longer GMC assembly process.


Much of the gas in the GMC is relatively diffuse, with extended regions seen only in $^{12}$CO(J=1--0). A brighter $^{13}$CO(J=1--0) ridge delineates the collision zone, but C$^{18}$O(J=1--0) emission is very weak, implying that the dense gas fraction -- and hence the star formation rate -- is low. Indeed, the 
bulk of the gas appears to be relatively quiescent, and evidence of active star formation is seen only in some outlying regions of the cloud. This is consistent with the suggestion that the cloud is still relatively young, with massive star-formation yet to commence in the vast majority of its volume. 

Despite its large mass, the GMC is not self-gravitating under the standard formulation of the Virial theorem. Instead, ram pressure from the colliding flows provides a promising candidate for an external confining pressure. This strongly suggests that self-gravity was not a necessary ingredient for its formation, and implies is that much of the molecular material (particularly the more diffuse gas) will disperse once the confining pressure is switched off -- a paradigmatic example of a transient, unbound GMC. 

Finally, we note possible evidence of fluid dynamical instabilities along the bottom ridge of the cloud, in the form of periodic drip-like formations in the molecular gas. These are likely best explained as Rayleigh-Taylor instabilities, possibly combined with other fluid-dynamical and dynamical instabilities such as the non-linear thin shell instability, which the numerical models suggest form readily throughout the swept-up gas.  

\acknowledgements
J. Dawson and E. Ntormousi wish to thank Mark Wardle and Shu-Ichiro Inutsuka for helpful discussions on the text, Rasmus Voss for provision of the OB data, and Patrick Hennebelle for simulation tips. This research has received funding from the European Research Council: E. Ntormousi is funded by the ERC Grant Agreement ``ORISTARS'' (no. 291294) and has received travel funding for this work from the ERC Grant no. 306483. J. Dawson received travel funding from a 2014 Scientific Mobility Grant from the French Embassy in Australia. K. Fierlinger was supported by the DFG cluster of excellence ``Origin and Structure of the Universe''. The NANTEN project was based on a mutual agreement between Nagoya University and the Carnegie Institute of Washington, and its operation was made possible thanks to contributions from companies and members of the Japanese public. We also extend our thanks to the past staff and students of Nagoya University who contributed to the observations utilized in this paper. The Australia Telescope Compact Array, Parkes and Mopra Telescopes are part of the Australia Telescope which is funded by the Commonwealth of Australia for operation as a National Facility managed by CSIRO. We acknowledge the use of the Southern H-Alpha Sky Survey Atlas (SHASSA), which is supported by the National Science Foundation. This publication also makes use of data products from the Wide-field Infrared Survey Explorer, which is a joint project of the University of California, Los Angeles, and the Jet Propulsion Laboratory/California Institute of Technology, funded by the National Aeronautics and Space Administration.

\bibliographystyle{apj}
\bibliography{gmcbib}

\begin{thebibliography}{78}
\expandafter\ifx\csname natexlab\endcsname\relax\def\natexlab#1{#1}\fi

\bibitem[{{Audit} \& {Hennebelle}(2005)}]{audit05}
{Audit}, E., \& {Hennebelle}, P. 2005, \aap, 433, 1

\bibitem[{{Ballesteros-Paredes} {et~al.}(1999){Ballesteros-Paredes},
  {Hartmann}, \& {V{\'a}zquez-Semadeni}}]{ballesteros99}
{Ballesteros-Paredes}, J., {Hartmann}, L., \& {V{\'a}zquez-Semadeni}, E. 1999,
  \apj, 527, 285

\bibitem[{{Bolatto} {et~al.}(2013){Bolatto}, {Wolfire}, \& {Leroy}}]{bolatto13}
{Bolatto}, A.~D., {Wolfire}, M., \& {Leroy}, A.~K. 2013, \araa, 51, 207

\bibitem[{{Bonnell} {et~al.}(2011){Bonnell}, {Smith}, {Clark}, \&
  {Bate}}]{bonnell11}
{Bonnell}, I.~A., {Smith}, R.~J., {Clark}, P.~C., \& {Bate}, M.~R. 2011,
  \mnras, 410, 2339

\bibitem[{{Bournaud} {et~al.}(2010){Bournaud}, {Elmegreen}, {Teyssier},
  {Block}, \& {Puerari}}]{bournaud10}
{Bournaud}, F., {Elmegreen}, B.~G., {Teyssier}, R., {Block}, D.~L., \&
  {Puerari}, I. 2010, \mnras, 409, 1088

\bibitem[{{Caswell}(1998)}]{caswell98}
{Caswell}, J.~L. 1998, \mnras, 297, 215

\bibitem[{{Clark} {et~al.}(2005){Clark}, {Bonnell}, {Zinnecker}, \&
  {Bate}}]{clark05}
{Clark}, P.~C., {Bonnell}, I.~A., {Zinnecker}, H., \& {Bate}, M.~R. 2005,
  \mnras, 359, 809

\bibitem[{{Dawson}(2013)}]{dawson13b}
{Dawson}, J.~R. 2013, \pasa, 30, 25

\bibitem[{{Dawson} {et~al.}(2008{\natexlab{a}}){Dawson}, {Kawamura}, {Mizuno},
  {Onishi}, \& {Fukui}}]{dawson08b}
{Dawson}, J.~R., {Kawamura}, A., {Mizuno}, N., {Onishi}, T., \& {Fukui}, Y.
  2008{\natexlab{a}}, \pasj, 60, 1297

\bibitem[{{Dawson} {et~al.}(2011{\natexlab{a}}){Dawson}, {McClure-Griffiths},
  {Dickey}, \& {Fukui}}]{dawson11b}
{Dawson}, J.~R., {McClure-Griffiths}, N.~M., {Dickey}, J.~M., \& {Fukui}, Y.
  2011{\natexlab{a}}, \apj, 741, 85

\bibitem[{{Dawson} {et~al.}(2011{\natexlab{b}}){Dawson}, {McClure-Griffiths},
  {Kawamura}, \& {Mizuno, N. et al.}}]{dawson11a}
{Dawson}, J.~R., {McClure-Griffiths}, N.~M., {Kawamura}, A., \& {Mizuno, N. et
  al.} 2011{\natexlab{b}}, \apj, 728, 127

\bibitem[{{Dawson} {et~al.}(2008{\natexlab{b}}){Dawson}, {Mizuno}, {Onishi},
  {McClure-Griffiths}, \& {Fukui}}]{dawson08a}
{Dawson}, J.~R., {Mizuno}, N., {Onishi}, T., {McClure-Griffiths}, N.~M., \&
  {Fukui}, Y. 2008{\natexlab{b}}, \mnras, 387, 31

\bibitem[{{de Avillez} \& {Berry}(2001)}]{avillez01}
{de Avillez}, M.~A., \& {Berry}, D.~L. 2001, \mnras, 328, 708

\bibitem[{{Dickman}(1978)}]{dickman78}
{Dickman}, R.~L. 1978, \apjs, 37, 407

\bibitem[{{Dobbs} \& {Bonnell}(2007)}]{dobbs07}
{Dobbs}, C.~L., \& {Bonnell}, I.~A. 2007, \mnras, 376, 1747

\bibitem[{{Dobbs} {et~al.}(2006){Dobbs}, {Bonnell}, \& {Pringle}}]{dobbs06}
{Dobbs}, C.~L., {Bonnell}, I.~A., \& {Pringle}, J.~E. 2006, \mnras, 371, 1663

\bibitem[{{Dobbs} {et~al.}(2011){Dobbs}, {Burkert}, \& {Pringle}}]{dobbs11b}
{Dobbs}, C.~L., {Burkert}, A., \& {Pringle}, J.~E. 2011, \mnras, 413, 2935

\bibitem[{{Elmegreen}(2000)}]{elmegreen00}
{Elmegreen}, B.~G. 2000, \apj, 530, 277

\bibitem[{{Elmegreen}(2007)}]{elmegreen07}
---. 2007, \apj, 668, 1064

\bibitem[{{Elmegreen}(2011)}]{elmegreen11}
---. 2011, \apj, 737, 10

\bibitem[{{Fierlinger} {et~al.}(2012){Fierlinger}, {Burkert}, {Diehl}, {Dobbs},
  {Hartmann}, {Krause}, {Ntormousi}, \& {Voss}}]{Fierlinger12}
{Fierlinger}, K.~M., {Burkert}, A., {Diehl}, R., {Dobbs}, C., {Hartmann},
  D.~H., {Krause}, M., {Ntormousi}, E., \& {Voss}, R. 2012, in Astronomical
  Society of the Pacific Conference Series, Vol. 453, Advances in Computational
  Astrophysics: Methods, Tools, and Outcome, ed. R.~{Capuzzo-Dolcetta},
  M.~{Limongi}, \& A.~{Tornamb{\`e}}, 25

\bibitem[{{Frerking} {et~al.}(1982){Frerking}, {Langer}, \&
  {Wilson}}]{frerking82}
{Frerking}, M.~A., {Langer}, W.~D., \& {Wilson}, R.~W. 1982, \apj, 262, 590

\bibitem[{{Fukui} {et~al.}(1999){Fukui}, {Onishi}, {Abe}, {Kawamura},
  {Tachihara}, {Yamaguchi}, {Mizuno}, \& {Ogawa}}]{fukui99}
{Fukui}, Y., {Onishi}, T., {Abe}, R., {Kawamura}, A., {Tachihara}, K.,
  {Yamaguchi}, R., {Mizuno}, A., \& {Ogawa}, H. 1999, \pasj, 51, 751

\bibitem[{{Garcia}(1994)}]{garcia94}
{Garcia}, B. 1994, \apj, 436, 705

\bibitem[{{Gaustad} {et~al.}(2001){Gaustad}, {McCullough}, {Rosing}, \& {Van
  Buren}}]{gaustad01}
{Gaustad}, J.~E., {McCullough}, P.~R., {Rosing}, W., \& {Van Buren}, D. 2001,
  \pasp, 113, 1326

\bibitem[{{Georgelin} {et~al.}(2000){Georgelin}, {Russeil}, {Amram},
  {Georgelin}, {Marcelin}, {Parker}, \& {Viale}}]{georgelin00}
{Georgelin}, Y.~M., {Russeil}, D., {Amram}, P., {Georgelin}, Y.~P., {Marcelin},
  M., {Parker}, Q.~A., \& {Viale}, A. 2000, \aap, 357, 308

\bibitem[{{Glover} \& {Clark}(2012)}]{glover12}
{Glover}, S.~C.~O., \& {Clark}, P.~C. 2012, \mnras, 421, 9

\bibitem[{{Glover} {et~al.}(2010){Glover}, {Federrath}, {Mac Low}, \&
  {Klessen}}]{glover10}
{Glover}, S.~C.~O., {Federrath}, C., {Mac Low}, M.-M., \& {Klessen}, R.~S.
  2010, \mnras, 404, 2

\bibitem[{{Green} {et~al.}(2012){Green}, {Caswell}, {Fuller}, {Avison},
  {Breen}, {Ellingsen}, {Gray}, {Pestalozzi}, {Quinn}, {Thompson}, \&
  {Voronkov}}]{green12}
{Green}, J.~A., {et~al.} 2012, \mnras, 420, 3108

\bibitem[{{Hartmann} {et~al.}(2001){Hartmann}, {Ballesteros-Paredes}, \&
  {Bergin}}]{hartmann01}
{Hartmann}, L., {Ballesteros-Paredes}, J., \& {Bergin}, E.~A. 2001, \apj, 562,
  852

\bibitem[{{Hartmann} {et~al.}(2012){Hartmann}, {Ballesteros-Paredes}, \&
  {Heitsch}}]{hartmann12}
{Hartmann}, L., {Ballesteros-Paredes}, J., \& {Heitsch}, F. 2012, \mnras, 420,
  1457

\bibitem[{{Haverkorn} {et~al.}(2006){Haverkorn}, {Gaensler},
  {McClure-Griffiths}, {Dickey}, \& {Green}}]{haverkorn06}
{Haverkorn}, M., {Gaensler}, B.~M., {McClure-Griffiths}, N.~M., {Dickey},
  J.~M., \& {Green}, A.~J. 2006, \apjs, 167, 230

\bibitem[{{Heiles}(1979)}]{heiles79}
{Heiles}, C. 1979, \apj, 229, 533

\bibitem[{{Heitsch} {et~al.}(2006){Heitsch}, {Slyz}, {Devriendt}, {Hartmann},
  \& {Burkert}}]{Heitsch_06}
{Heitsch}, F., {Slyz}, A.~D., {Devriendt}, J.~E.~G., {Hartmann}, L.~W., \&
  {Burkert}, A. 2006, \apj, 648, 1052

\bibitem[{{Heitsch} {et~al.}(2009){Heitsch}, {Stone}, \&
  {Hartmann}}]{heitsch09}
{Heitsch}, F., {Stone}, J.~M., \& {Hartmann}, L.~W. 2009, \apj, 695, 248

\bibitem[{{Henley} {et~al.}(2010){Henley}, {Shelton}, {Kwak}, {Joung}, \& {Mac
  Low}}]{henley10}
{Henley}, D.~B., {Shelton}, R.~L., {Kwak}, K., {Joung}, M.~R., \& {Mac Low},
  M.-M. 2010, \apj, 723, 935

\bibitem[{{Hennebelle} \& {P{\'e}rault}(1999)}]{hennebelle99}
{Hennebelle}, P., \& {P{\'e}rault}, M. 1999, \aap, 351, 309

\bibitem[{{Heyer} {et~al.}(2001){Heyer}, {Carpenter}, \& {Snell}}]{heyer01}
{Heyer}, M.~H., {Carpenter}, J.~M., \& {Snell}, R.~L. 2001, \apj, 551, 852

\bibitem[{{Hill} {et~al.}(2012){Hill}, {Joung}, {Mac Low}, {Benjamin},
  {Haffner}, {Klingenberg}, \& {Waagan}}]{hill12}
{Hill}, A.~S., {Joung}, M.~R., {Mac Low}, M.-M., {Benjamin}, R.~A., {Haffner},
  L.~M., {Klingenberg}, C., \& {Waagan}, K. 2012, \apj, 750, 104

\bibitem[{{Inoue} \& {Inutsuka}(2008)}]{inoue08}
{Inoue}, T., \& {Inutsuka}, S. 2008, \apj, 687, 303

\bibitem[{{Inoue} \& {Inutsuka}(2009)}]{inoue09}
{Inoue}, T., \& {Inutsuka}, S.-i. 2009, \apj, 704, 161

\bibitem[{{Inoue} \& {Inutsuka}(2012)}]{inoue12}
---. 2012, \apj, 759, 35

\bibitem[{{Inutsuka} {et~al.}(2014){Inutsuka}, {Inoue}, \&
  {Iwasaki}}]{inutsuka14}
{Inutsuka}, S.-I., {Inoue}, T., \& {Iwasaki}, K. 2014, \apj, submitted

\bibitem[{{Kaltcheva} \& {Scorcio}(2010)}]{kaltcheva10}
{Kaltcheva}, N., \& {Scorcio}, M. 2010, \aap, 514, A59

\bibitem[{{Kaltcheva}(1998)}]{kaltcheva98}
{Kaltcheva}, N.~T. 1998, \aaps, 128, 309

\bibitem[{{Kim} \& {Ostriker}(2006)}]{kim06}
{Kim}, W.-T., \& {Ostriker}, E.~C. 2006, \apj, 646, 213

\bibitem[{{Kim} {et~al.}(2002){Kim}, {Ostriker}, \& {Stone}}]{kim02}
{Kim}, W.-T., {Ostriker}, E.~C., \& {Stone}, J.~M. 2002, \apj, 581, 1080

\bibitem[{{Koyama} \& {Inutsuka}(2000)}]{koyama00}
{Koyama}, H., \& {Inutsuka}, S.-I. 2000, \apj, 532, 980

\bibitem[{{Kutner} \& {Ulich}(1981)}]{kutner81}
{Kutner}, M.~L., \& {Ulich}, B.~L. 1981, \apj, 250, 341

\bibitem[{{Larson}(1981)}]{larson81}
{Larson}, R.~B. 1981, \mnras, 194, 809

\bibitem[{{Mac Low}(1999)}]{MacLow_1999}
{Mac Low}, M.-M. 1999, \apj, 524, 169

\bibitem[{{Maloney}(1990)}]{maloney90}
{Maloney}, P. 1990, \apjl, 348, L9

\bibitem[{{Matsunaga}(2002)}]{matsunagaphd}
{Matsunaga}, K. 2002, PhD thesis, Nagoya University

\bibitem[{{Mauch} {et~al.}(2003){Mauch}, {Murphy}, {Buttery}, {Curran},
  {Hunstead}, {Piestrzynski}, {Robertson}, \& {Sadler}}]{mauch03}
{Mauch}, T., {Murphy}, T., {Buttery}, H.~J., {Curran}, J., {Hunstead}, R.~W.,
  {Piestrzynski}, B., {Robertson}, J.~G., \& {Sadler}, E.~M. 2003, \mnras, 342,
  1117

\bibitem[{{McClure-Griffiths} {et~al.}(2002){McClure-Griffiths}, {Dickey},
  {Gaensler}, \& {Green}}]{mcclure02}
{McClure-Griffiths}, N.~M., {Dickey}, J.~M., {Gaensler}, B.~M., \& {Green},
  A.~J. 2002, \apj, 578, 176

\bibitem[{{McClure-Griffiths} {et~al.}(2003){McClure-Griffiths}, {Dickey},
  {Gaensler}, \& {Green}}]{mcclure03}
---. 2003, \apj, 594, 833

\bibitem[{{McClure-Griffiths} {et~al.}(2006){McClure-Griffiths}, {Ford},
  {Pisano}, {Gibson}, {Staveley-Smith}, {Calabretta}, {Dedes}, \&
  {Kalberla}}]{mcclure06}
{McClure-Griffiths}, N.~M., {Ford}, A., {Pisano}, D.~J., {Gibson}, B.~K.,
  {Staveley-Smith}, L., {Calabretta}, M.~R., {Dedes}, L., \& {Kalberla},
  P.~M.~W. 2006, \apj, 638, 196

\bibitem[{{McCray} \& {Kafatos}(1987)}]{mccray87}
{McCray}, R., \& {Kafatos}, M. 1987, \apj, 317, 190

\bibitem[{{McLeod} \& {Whitworth}(2013)}]{McLeod_Whitworth_2013}
{McLeod}, A.~D., \& {Whitworth}, A.~P. 2013, \mnras, 431, 710

\bibitem[{{Mizuno} \& {Fukui}(2004)}]{mizuno04}
{Mizuno}, A., \& {Fukui}, Y. 2004, in Astronomical Society of the Pacific
  Conference Series, Vol. 317, Milky Way Surveys: The Structure and Evolution
  of our Galaxy, ed. D.~{Clemens}, R.~{Shah}, \& T.~{Brainerd}, 59

\bibitem[{{Ntormousi} {et~al.}(2011){Ntormousi}, {Burkert}, {Fierlinger}, \&
  {Heitsch}}]{ntormousi11}
{Ntormousi}, E., {Burkert}, A., {Fierlinger}, K., \& {Heitsch}, F. 2011, \apj,
  731, 13

\bibitem[{{Polk} {et~al.}(1988){Polk}, {Knapp}, {Stark}, \& {Wilson}}]{polk88}
{Polk}, K.~S., {Knapp}, G.~R., {Stark}, A.~A., \& {Wilson}, R.~W. 1988, \apj,
  332, 432

\bibitem[{{Rizzo} \& {Arnal}(1998)}]{rizzo98}
{Rizzo}, J.~R., \& {Arnal}, E.~M. 1998, \aap, 332, 1025

\bibitem[{{Solomon} {et~al.}(1987){Solomon}, {Rivolo}, {Barrett}, \&
  {Yahil}}]{solomon87}
{Solomon}, P.~M., {Rivolo}, A.~R., {Barrett}, J., \& {Yahil}, A. 1987, \apj,
  319, 730

\bibitem[{{Stone} \& {Gardiner}(2007)}]{stone07}
{Stone}, J.~M., \& {Gardiner}, T. 2007, \apj, 671, 1726

\bibitem[{{Tasker} \& {Tan}(2009)}]{tasker09}
{Tasker}, E.~J., \& {Tan}, J.~C. 2009, \apj, 700, 358

\bibitem[{{Teyssier}(2002)}]{Teyssier02}
{Teyssier}, R. 2002, \aap, 385, 337

\bibitem[{{V{\'a}zquez-Semadeni}(2010)}]{vazquez10}
{V{\'a}zquez-Semadeni}, E. 2010, in Astronomical Society of the Pacific
  Conference Series, Vol. 438, Astronomical Society of the Pacific Conference
  Series, ed. R.~{Kothes}, T.~L. {Landecker}, \& A.~G. {Willis}, 83

\bibitem[{{V{\'a}zquez-Semadeni} {et~al.}(2011){V{\'a}zquez-Semadeni},
  {Banerjee}, {G{\'o}mez}, {Hennebelle}, {Duffin}, \& {Klessen}}]{vazquez11}
{V{\'a}zquez-Semadeni}, E., {Banerjee}, R., {G{\'o}mez}, G.~C., {Hennebelle},
  P., {Duffin}, D., \& {Klessen}, R.~S. 2011, \mnras, 414, 2511

\bibitem[{{V{\'a}zquez-Semadeni} {et~al.}(2006){V{\'a}zquez-Semadeni}, {Ryu},
  {Passot}, {Gonz{\'a}lez}, \& {Gazol}}]{vazquez06}
{V{\'a}zquez-Semadeni}, E., {Ryu}, D., {Passot}, T., {Gonz{\'a}lez}, R.~F., \&
  {Gazol}, A. 2006, \apj, 643, 245

\bibitem[{{Vishniac}(1983)}]{Vishniac_83}
{Vishniac}, E.~T. 1983, \apj, 274, 152

\bibitem[{{Vishniac}(1994)}]{Vishniac_94}
---. 1994, \apj, 428, 186

\bibitem[{{Voss} {et~al.}(2009){Voss}, {Diehl}, {Hartmann}, {Cervi{\~n}o},
  {Vink}, {Meynet}, {Limongi}, \& {Chieffi}}]{Voss09}
{Voss}, R., {Diehl}, R., {Hartmann}, D.~H., {Cervi{\~n}o}, M., {Vink}, J.~S.,
  {Meynet}, G., {Limongi}, M., \& {Chieffi}, A. 2009, \aap, 504, 531

\bibitem[{{Wada} {et~al.}(2000){Wada}, {Spaans}, \& {Kim}}]{wada00}
{Wada}, K., {Spaans}, M., \& {Kim}, S. 2000, \apj, 540, 797

\bibitem[{{Weaver} {et~al.}(1977){Weaver}, {McCray}, {Castor}, {Shapiro}, \&
  {Moore}}]{Weaver_77}
{Weaver}, R., {McCray}, R., {Castor}, J., {Shapiro}, P., \& {Moore}, R. 1977,
  \apj, 218, 377

\bibitem[{{Wolfire} {et~al.}(1995){Wolfire}, {Hollenbach}, {McKee}, {Tielens},
  \& {Bakes}}]{Wolfire95}
{Wolfire}, M.~G., {Hollenbach}, D., {McKee}, C.~F., {Tielens}, A.~G.~G.~M., \&
  {Bakes}, E.~L.~O. 1995, \apj, 443, 152

\bibitem[{{Wright} {et~al.}(2010){Wright}, {Eisenhardt}, {Mainzer}, {Ressler},
  {Cutri}, {Jarrett}, {Kirkpatrick}, {Padgett}, {McMillan}, {Skrutskie},
  {Stanford}, {Cohen}, {Walker}, {Mather}, {Leisawitz}, {Gautier}, {McLean},
  {Benford}, {Lonsdale}, {Blain}, {Mendez}, {Irace}, {Duval}, {Liu}, {Royer},
  {Heinrichsen}, {Howard}, {Shannon}, {Kendall}, {Walsh}, {Larsen}, {Cardon},
  {Schick}, {Schwalm}, {Abid}, {Fabinsky}, {Naes}, \& {Tsai}}]{wright10}
{Wright}, E.~L., {et~al.} 2010, \aj, 140, 1868

\bibitem[{{Yonekura} {et~al.}(2005){Yonekura}, {Asayama}, {Kimura}, {Ogawa},
  {Kanai}, {Yamaguchi}, {Barnes}, \& {Fukui}}]{yonekura05}
{Yonekura}, Y., {Asayama}, S., {Kimura}, K., {Ogawa}, H., {Kanai}, Y.,
  {Yamaguchi}, N., {Barnes}, P.~J., \& {Fukui}, Y. 2005, \apj, 634, 476

\end{thebibliography}

\end{document}